\documentclass[11pt]{article}
\usepackage{amsmath, amssymb, amsthm,latexsym,color,times}
\usepackage{tikz,pgf}
\usepackage[pdftex]{hyperref}
\usepackage{fullpage}
\usepackage{subfigure}
\usepackage{enumerate}
\usepackage{framed}
\usepackage{algorithm}
\usepackage{algorithmic}

\def\A {\mathcal{A}}

\def \SA {\mathcal{SA}}
\def \LA {\mathcal{SA}_{x_0}}

\newtheorem{theorem}{Theorem}
\newtheorem{lemma}{Lemma}

\newtheorem{corollary}{Corollary}
\newtheorem{definition}{Definition}

\def\E{\mathbb{E}}

\def\twocolfig#1#2#3#4{
\begin{figure}[ht]
\begin{center}
\includegraphics[width=#1in]{#2}
\end{center}
\caption{#3}
\label{#4}
\end{figure}
}

\makeatletter
\let\@fnsymbol\@arabic 

\begin{document}

\setcounter{page}{0}

\title{A Generalization of Multiple Choice Balls-into-Bins: Tight Bounds}

\date{}

\author{Gahyun Park \\
        School of Mathematical Sciences\\
        Rochester Institute of Technology \\
        Rochester, NY 14623\\
        gxpsma@rit.edu }

\maketitle

\begin{abstract}
This paper investigates a general version of the multiple choice model called the $(k,d)$-choice process in which $n$ balls are assigned to $n$ bins. In the process, $k<d$ balls are placed  into the $k$ least loaded  out of $d$ bins chosen  independently and uniformly at random in each of $\frac{n}{k}$ rounds. The primary goal is to derive tight bounds on the maximum bin load for $(k,d)$-choice for any $1 \leq k < d \leq n$.  Our results enable one to choose suitable parameters $k$ and $d$ for which the $(k,d)$-choice process achieves the optimal tradeoff between the maximum bin load and message cost: a constant maximum load and $O(n)$ messages. The maximum load for a heavily loaded case where $m>n$ balls are placed into $n$ bins is also presented 
for the case $d \geq 2k$. Potential applications are discussed such as distributed storage as well as parallel job scheduling in a cluster.
 \end{abstract}


\normalsize

\section{Introduction}
In the classical single choice balls-into-bins problem, a ball is placed into a bin chosen independently and uniformly at random ({\it i.u.r.}). It is common knowledge that the maximum bin load of this basic process after $n$ balls are placed into $n$ bins is $(1+o(1))(\ln n/\ln \ln n)$ with high probability (w.h.p.) \cite{RS98}. In the multiple choice paradigm,  each ball is placed into the least loaded out of $d\geq 2$ bins chosen {\it i.u.r.}   Azar et al. showed that the maximum load in this case is exponentially reduced to $\ln\ln n/\ln d + O(1)$ w.h.p. \cite{ABKU94}.  Since then, numerous variations of the standard multiple choice problem have been investigated (e.g. \cite{Power96, V99, Talwar07,G08, PTW10, Cloud13}). 
For example, Berenbrink et al.\cite{BCSV00} and Talwar et al. \cite{Talwar13}  proved  that the gap between the maximum and average load still remains $\ln\ln n/\ln d + O(1)$ even if the number of balls grows  unboundedly large.  Czumaj and Stemann \cite{Czumaj} proposed an adaptive algorithm (i.e. the number of choices made by each ball varies depending on the load of the chosen bins) that achieves $O(\ln\ln n/\ln d)$ maximum load with  $(1+o(1))n$ message cost. \footnote{ The message cost is the cost of network communication incurred by bin probing and defined by the number of bins to be probed.}
Parallel versions\cite{Adler95, Stemann96,Power96, Lenzen, BCEFN12}  of balanced allocation 
have been studied. Recent works addressed near optimal adaptive algorithms: a constant maximum bin load using an average of $O(1)$ bin choices per ball \cite{Lenzen, BSS13}.

In our prior work \cite{PARK11}\footnote{A preliminary version of this paper was  published in PODC' 11, pages 297-298, titled with
``Brief announcement: A Generalization of multiple choice balls-into-bins".}, we posed the following questions: If we place $2$ balls at a time into $2$ least loaded out of $3$ bins chosen i.u.r., is the maximum load  still $\Theta(\ln\ln n)$?. What would occur if $5$ balls are placed into $8$ bins or $99$ balls into $100$ bins? In general, if $k<d$ balls  are assigned to the $k$ least loaded  among $d$ possible destinations chosen i.u.r., which we call $(k,d)$-choice, what is the maximum load of any bin after all $n$ balls are placed into $n$ bins? 

In this paper,  we derive tight bounds on the maximum load of $(k,d)$-choice as a function of three parameters $k,d,$ and $n$, which in turn allow one to choose appropriate values of $k$ and $d$ to achieve a striking balance between the maximum bin load and message cost: a constant maximum load with  $2n$ messages, or $o(\ln\ln n)$ maximum load with $(1+o(1))n$ messages. This suggests that our non-adaptive allocation scheme is  near-optimal (for appropriate $k$ and $d$), and outperforms existing non-adaptive allocation schemes; to the best of our knowledge, none of previously known algorithms using  $O(n)$ messages achieve a constant maximum load.\footnote{A balls-into-bins model is called non-adaptive if the number of choices per ball is fixed.}

The $(k,d)$-choice model represents the full spectrum of balanced allocations that lie between the single- and multi-choice algorithms-- for small $k$, $(k,d)$-choice acts like the standard $d$-choice, while it converges to the classic single choice balls-into-bins problem for $k \approx d$ and large $d$. 
Our work is similar in spirit to the $(1+\beta)$-choice algorithm proposed by Peres et al \cite{PTW10}, where each ball goes to the lesser loaded of two random bins with probability $\beta$ and a random bin with probability $1-\beta$, in that both schemes can be viewed as  a mix between single- and multiple-choice strategies, though these two models exhibit no other structural similarities. The $(k,d)$-choice algorithm is a (semi) parallel version of the basic sequential $d$-choice, but fundamentally different from existing parallel balanced allocations\cite{Adler95, Stemann96,Power96, Lenzen, BCEFN12}: In previous parallel models, each ball carries out bin probing independently from other balls, whereas, in our case,  a group of  $k$ balls shares information on bin state and uses the information to lower the maximum load. From a technical point of view, Markov chain coupling and layered induction \cite{ABKU94, BCSV00} is the inspiration for our analysis. We extend the existing analysis to a general setting where there are strong data dependencies arising from the behavior of $k$ balls.

We note that there are some ambiguities in the $(k,d)$-choice allocation policy. For example, suppose that  four bins -- bin1,..., bin4-- contain $3, 2, 1$, and $0$ balls, respectively,  in the beginning of a round for the $(3,4)$-choice process.
We consider three different scenarios:
\begin{enumerate}[(a)]
\item Each of the four bins is sampled once.
\item Each of bin2 and bin3 is sampled once, and bin4 is sampled twice.
\item Each of bin1 and bin4 is sampled twice.
\end{enumerate}
In the first scenario, each of bin2, bin3, and bin 4 receives a new ball; however, in the next two cases, some bins are sampled multiple times, creating ambiguities on the destinations of the three balls. In  scenario (b), one option is to assign each ball to each sampled bin, and another option is to assign two balls to bin4 and the other to bin3. The last scenario  is even more problematic since only two destinations are available. 

We eliminate this ambiguity by imposing the restriction on the allocation policy; bins sampled $m \geq 1$ times can receive at most  $m$ balls.  This rule can be explained in a different way: In each round, each of $d$ balls (instead of $k$ balls) is placed sequentially into a random bin. At the end of the round,  $d-k$ balls (among the $d$ balls belonging to the round) with maximal \emph{height} are removed, where the \emph{height} of a ball is defined as the number of balls in the bin containing the ball right after it is placed. According to the policy,  bin3 receives a ball and bin4 receives two balls in scenario (b), and bin1 receives one ball and bin4 receives two in scenario (c). Note that 
 the $(k,d)$-choice policy is not always optimal; a better strategy is to assign one ball to bin3 and two balls to bin4 in scenario (a),  and all three balls to bin4 in scenario (c).  In practice, one can easily modify the policy to increase load balance.

The rest of the paper is organized as follows. In the rest of this section, we discuss our main results,  simulation results, and potential applications. Section~\ref{sec:model} presents the model, and definitions  and a list of notations used throughout this paper. We present some key properties of $(k,d)$-choice in Section~\ref{sec:properties} and analyze the upper and lower bounds on the maximum loads in Section~\ref{sec:up} and Section~\ref{sec:lo}, respectively. We provide proofs of some lemmas  in Section~\ref{sec:lemmas}, and conclude the paper in Section~\ref{sec:con}.

\subsection{Main Results}\label{sec:results}
We assume that $n$ is a multiple of $k$ and the $(k,d)$-choice process consists of $n/k$ rounds. Our balls-into-bins model is described as follows.
\begin{framed}
{\bf The $(k,d)$-choice process:}\\
In each round,  $k < d$ balls are placed into the $k$ least loaded (with ties broken randomly) out of $d$ bins chosen i.u.r. (with replacement) such that
a bin sampled $m \geq 1$ times receives at most $m$ balls.
\end{framed}
Our main result is formalized as follows. 
\begin{theorem} \label{thm:1}
For $1 \leq k < d \leq n$, let 
$d_k = \frac{d}{d-k}$. 
Let $M(k,d, n)$ denote the maximum load after $n$ balls are placed into $n$ bins under  the $(k,d)$-choice process.  
If $n \to \infty$, the following holds with probability $1 - o\left(1\right )$. \\
(i) If $ d_k  =  O(1)$, then 
\begin{align}
\frac{\ln \ln n}{\ln (d-k+1)} - O(1) \leq M(k,d,n) \leq   \frac{\ln \ln n}{\ln (d-k+1)} + O(1). \label{eq:main1}
\end{align}
(ii) If $d_k \to \infty$ as $n \to \infty$,   then
\begin{align}
 \frac{\ln\ln n}{\ln (d-k+1)} + (1-o(1)) \frac{\ln d_k}{\ln\ln d_k} \leq M(k,d,n)
\leq  \frac{\ln\ln n}{\ln (d-k+1)} + (1+o(1)) \frac{\ln d_k}{\ln\ln d_k}. \label{eq:main2}
\end{align}
\end{theorem}
As $d_k \geq e^{(\ln\ln n)^3}$, we have $\frac{\ln d_k}{\ln\ln d_k} \gg \ln\ln n$, and hence (\ref{eq:main2}) can be simplified as follows.
\begin{corollary}\label{cor:1}
If $d_k = \frac{d}{d-k} \geq  e^{(\ln\ln n)^3}$, then with probability $1-o\left(1\right)$
\begin{align}
(1-o(1)) \frac{\ln d_k}{\ln\ln d_k} \leq M(k,d,n) \leq (1+o(1)) \frac{\ln d_k}{\ln\ln d_k}. \label{eq:main3}
\end{align}
\end{corollary}
We discuss several interesting consequences derived from the main result.
If we choose the smallest $k$ ($= 1$) and hence $d_k = O(1)$, the result (\ref{eq:main1}) is reduced to the maximum load for the standard $d$-choice algorithm.  At another extreme, if $k \approx d$ for large $d \approx n$ then  (\ref{eq:main3}) implies that the maximum load becomes $(1+o(1))(\ln n/\ln\ln n)$; this agrees with the maximum load for the classical single choice algorithm. The true benefit of the $(k,d)$-choice scheme lies between these two extremes. For example, if $k \geq \Theta(\ln^2 n)$ and $d-k = \Theta(\ln n)$, then the result (\ref{eq:main2}) implies that $(k,d)$-choice achieves $o(\ln\ln n)$ maximum load  using the asymptotically minimal cost $(1+o(1))n$ of messages.  To our best knowledge, the previously  known result using $(1+o(1))n$ messages is an adaptive algorithm with $O(\ln\ln n)$ maximum load presented in \cite{Czumaj}. Another example is that if $d-k+1 \geq \Omega(\ln n)$ and $d_k = O(1)$ (such as $k = \Theta(\text{polylog } n)$ and $d = 2k$),  then (\ref{eq:main1}) suggests that a constant maximum load is achieved at the cost of $O(n)$ messages, which is comparable to the best known result of an adaptive algorithm in \cite{Lenzen}.  This suggests that our non-adaptive allocation scheme  performs  as well as the best known adaptive algorithm.

We obtain the following (partial) results on the heavily loaded case where the number of balls exceeds the number of bins.
\begin{theorem}   \label{thm:2}  Let $M(k,d,m,n)$ denote the maximum bin load after $m > n$ balls are placed into $n$ bins following the $(k,d)$-choice process. If $d \geq 2k$, then the maximum load is
\begin{align}
\frac{\ln\ln n}{\ln (d-k+1)} - O(1) \leq M(k,d,m,n) \leq \frac{\ln\ln n}{\ln \lfloor d/k \rfloor} + O(1), \label{eq:heavy}
\end{align}
with probability $1-o(1/n)$.
\end{theorem}

\subsection{Experimental Results}
In Table 1, we present simulation results of $(k,d)$-choice after $n$ balls are placed into $n$ bins using $n = 3\cdot 2^{16}$ and varying $k$ and $d$ values. A pseudo random number generator is used to sample $d$ random bins in each round of the process. The maxim load shown in the table  is obtained after running the simulation ten times in each. All $k$ values we have chosen divide $n$ so that exactly $k$ balls are inserted in each round.  In the second and third columns, the maximum load of single-choice and two-choice is given. It is worth of note that the result of $(8,9)$-choice is close to that of two-choice and $(128, 193)$-choice outperforms two-choice and achieves the same maximum load $2$ as $(1, 193)$-choice. We also remark that $(64,65)$-choice performs noticeably better than single-choice.
\begin{center}
\begin{table}
\caption{ The maximum bin load for $(k,d)$-choice with $n = 3\cdot2^{16}$ and varying $k$ and $d$ values}
\begin{tabular}{| l |  r | r | r | r | r | r | r  |r  |r | r  |}
\hline                        
$\;$ & $d=1$ & $d=2$ & $d=3$ & $d=5$ & $d = 9$ &  $d = 17$ & $d = 25$  & $d = 49$ & $d = 65$  & $d = 193$ \\
\hline
$k=1$  & $7,8,9$ & $3,4$ & $3$ & $2$ & $2$ &  $2$ & $2$ & $2$ & $2$ &  $2$\\
$k=2$  & $-$ & $-$ & $4$ & $3$ & $2$ & $2$ & $2$  & $2$ & $2$  &  $2$\\
$k=3$  & $-$ & $-$ & $-$ & $3$ & $2$ &  $2$ & $2$ & $2$ & $2$ &  $2$\\
$k=4$   & $-$ & $-$ & $-$ & $4$ & $3$ & $2$ & $2$  & $2$ & $2$ &  $2$\\
$k=6$   & $-$ & $-$ & $-$ & $-$ & $3$ &  $2$ & $2$  & $2$ & $2$ &  $2$ \\
$k=8$   & $-$ & $-$ & $-$ & $-$ & $4$ &  $2, 3$ & $2$  & $2$ & $2$  &  $2$ \\
$k=12$  & $-$  & $-$ & $-$ & $-$ & $-$  & $3$ & $2$  & $2$ & $2$  &  $2$ \\
$k=16$  & $-$  & $-$ & $-$ & $-$ & $-$ &  $4, 5$ & $3$  & $2$ & $2$  &  $2$\\
$k=24$  & $-$  & $-$ & $-$ & $-$ & $-$ &  $-$ & $5$  & $2$ & $2$  &  $2$\\
$k=32$   & $-$ & $-$ & $-$ & $-$ & $-$ &  $-$ & $-$  & $3$ & $2$  &  $2$\\
$k=48$   & $-$ & $-$ & $-$ & $-$ & $-$ &  $-$ & $-$  & $5$ & $3$  &  $2$\\
$k=64$   & $-$ & $-$ & $-$ & $-$ & $-$ &  $-$ & $-$  & $-$ & $5$  & $2$\\
$k=96$   & $-$ & $-$ & $-$ & $-$ & $-$ &  $-$ & $-$  & $-$ & $-$  & $2$\\
$k=128$   & $-$ & $-$ & $-$ & $-$ & $-$ &  $-$ & $-$  & $-$ & $-$  &  $2$\\
$k=192$   & $-$ & $-$ & $-$ & $-$ & $-$ &  $-$ & $-$  & $-$ & $-$  & $5,6$\\
\hline 
\end{tabular}
\end{table}
\end{center}
\subsection{Applications}\label{sec:app}
The $(k,d)$-choice allocation scheme is used for a parallel job scheduling in a cluster environment \cite{SPARROW} as $(k,d$)-choice enables low response time. 
Suppose that a job consists of $k$ tasks to be scheduled in parallel, and each task issues $d$ random probes individually (as in $d$-choice). In this case,  it is likely that there will be  a ball/task whose $d$ possible destinations are all heavily loaded. Since a job's completion time is determined by the task finishing last, the performance of the standard multiple choice degrades as a job's parallelism increases. Our $(k,d)$-choice model solves this problem by letting $k$ tasks share information across all the probes in a job, which effectively reduces the chance for any tasks to commit to a heavily loaded worker machine.

A distributed storage system is another application domain to which $(k,d)$-choice is naturally applicable. Data replication and fragmentation are widely used in this setting to increase file availability, fault tolerance, and load balance. Suppose that a new file is created and replicated into $k$ copies (or that a large file is split into $k$ chunks), and each of the replicas (or chunks) is to be stored on servers. The $(k,d)$-choice scheme provides a simple and efficient solution for fast allocation and  load balance with the minimum message cost; $k$ replicas (or chunks) are stored on the $k$ least loaded  out of $d$ servers chosen randomly.  
If, for example, $d = k+1$ and $k=\Theta(\ln n)$, then $(k,d)$-choice provides the asymptotically same maximum load as that of the two-choice scheme at the half of the message cost of two-choice.
In case of data partitioning, if a file search requires retrieving all $k$ chunks of the file, then the search operation costs $k+1$, which is (asymptotically) minimum and approximately half of the search cost for two-choice.

\section{Model, Notations, and Definitions}\label{sec:model}
\subsection{The Model and Notations}\label{sec:notation}
We will assume that, at the end of each round, bins are sorted in decreasing order in terms of bin load (with ties broken randomly). By bin $x$, we denote the $x$th most loaded bin (at the time of consideration); that is, bin $1$ is the most loaded bin, bin $2$ is the second most loaded bin, and so on.  Then the $(k,d)$-choice process can be viewed as Markovian with the state space composed of the sorted bin load vectors. 

The {\it height} of a ball is the number of balls in the bin containing the ball immediately after it is placed. In case the two balls fall in the same bin in the same round, each of them can be assumed to have a different height. For the purpose of analysis, we may assume that one of the ball is placed first and in turn has less height than the other.
The following is the list of notations used in this paper.
\begin{itemize}
\item $\A = \A(k,d)$ is the $(k,d)$-choice algorithm.
\item $\SA = \SA(k, k)$ is the classical single choice equivalence, where $k$ balls are placed into $k$ bins in each round.
\item
$B_x^{\A}(r)$ is the number of balls in bin $x$ at the end of the $r$th round resulting from algorithm $\A$. 
\item
$B_{\leq x}^{\A}(r) = B_1^{\A}(r) + \ldots + B_x^{\A}(r)$ is the number of balls in the $x$ most loaded bins at the end of the $r$th round resulting from algorithm $\A$. 
\item
$\mu_y^{\A}(r)$ is the number of balls with height at least $y$ at the end of the $r$th round resulting from algorithm $\A$.
\item
$\nu_y^{\A}(r)$ is the number of bins with at least $y$ balls at the end of the $r$th round resulting from algorithm $\A$.
\item
$M(k,d,n)$ denotes the maximum load after $n$ balls are placed into bins resulting from the $(k,d)$-choice process.
\item
$\delta = \delta(n) = \frac{\ln\ln\ln n}{\ln\ln n}$.
\item
$d_k = \frac{d}{d-k}$
\item 
$\text{polylog } n = (\ln n)^{\theta(1)}$
\end{itemize}

For simplicity, we sometimes use $B_x^{\A}$ (or $B_{x}$) to denote $B_x^{\A}(n/k)$, the number of balls in bin $x$ at the end of the $(k,d)$-choice process. We also use  $\nu_y$  and $\mu_y$ to denote $\nu_{y}^{\A}(n/k)$ and $\mu_y^{\A}(n/k)$, respectively.

\subsection{Definitions}
The $(k,d)$-choice process can run sequentially.  In some part of analysis, we treat $(k,d)$-choice as a sequential process and need a  notion of bin state at any time $1\leq t \leq n$. 
\begin{definition} {\bf Serialization of $(k,d)$-choice}
For $1\leq r \leq n/k$, let $\sigma_{r} = (i_{r, 1}, i_{r, 2}, \ldots, i_{r, k})$ be a permutation of $\{1, 2, \ldots, k\}$. 
In each round $r$, a set $S_r$ of $d$ bins is chosen i.u.r. and each of $k$ balls is placed sequentially into a bin as follows.
The first ball falls into the $i_{r, 1}$th least loaded bin in $S_r$, the second ball falls into the $i_{r, 2}$th least loaded bin in $S_r$, and so on.
Let $\sigma = (\sigma_1, \sigma_2, \ldots, \sigma_{n/k})$. By $\A_{\sigma} = \A_{\sigma}(k,d)$, we denote the serialized version of $(k,d)$-choice induced by $\sigma$.   
\end{definition}

\begin{itemize}
\item
$B_x^{\A_{\sigma}}(t)$ is the number of balls in bin $x$ at time $t$ (right after the $t$th ball is placed and the bins are sorted), $1\leq t\leq n$, resulting from $\A_{\sigma}$.
\item $p_x^{\A{_\sigma}}(t)$ is the probability that the $t$th ball, $1\leq t \leq n$,  is placed into bin $x$ resulting from algorithm $\A_{\sigma}$, where bin $x$ is the $x$th most loaded bin at time $t-1$ (i.e., right before the $t$th ball is placed).
\end{itemize}
Let $\sigma = (\sigma_1, \ldots, \sigma_{n/k})$ and $\pi = (\pi_1, \ldots, \pi_{n/k})$, where $\sigma_r$ and $\pi_r$ are permutations of $\{1, 2, \ldots, k\}$. If  $\sigma_r \neq \pi_r$ for some $r$, then
$B_x^{\A_{\sigma}}(t)$ and $B_x^{\A_{\pi}}(t)$ may have different probability distributions and 
$p_x^{\A_{\sigma}}(t) \neq p_x^{\A_{\pi}}(t)$ in general.
There should be no confusion between $B_{x}^\A(r)$ and $B_{x}^{\A_{\sigma}}(t)$: The former is the load of bin $x$ at the end of $r$th round (right after $rk$ balls are placed), while the latter is the load of bin $x$ at time $t$ (right after $t$ balls are placed). 
We use $B_x^{\A_{\sigma}}$ to denote $B_{x}^{\A_{\sigma}}(n)$.

\begin{definition}  \label{def:maj}
Let $\A_1$ and $\A_2$ be allocation processes starting with $n$ empty bins.
Let $B_{\leq x}^{\A_i}  = B_1^{\A_i}+B_2^{\A_i}+ \ldots +B_x^{\A_i}$ represent the number of balls in the $x$ most loaded 
bins at the end of process $\A_i$, $i=1, 2$, where $B_x^{\A_i}$ denotes the number of balls in the $x$th most loaded bin. 
\\ 
i) We say that $\A_1$ and $\A_2$ are {\it equivalent}, denoted  $\A_1 \equiv \A_2$, if 
$$
\Pr\left( B_{\leq x}^{\A_1}  \geq t \right) = \Pr\left( B_{\leq x}^{\A_2} \geq t \right), 
$$
for $1\leq x \leq n$ and $t \geq 0$.\\
ii) We say that $\A_1$ is {\it majorized} by $\A_2$, denoted  $\A_1 \leq_{mj} \A_2$, if 
\begin{align}
\Pr\left( B_{\leq x}^{\A_1}  \geq t \right) \leq \Pr\left( B_{\leq x}^{\A_2} \geq t \right),
\end{align}
for $1\leq x \leq n$ and $t \geq 0$.\\
iii) We say that $\A_1$ is {\it dominated} by $\A_2$, denoted
$\A_1 \leq_{dm} \A_2$, 
if 
\begin{equation}
\Pr \left(B_x^{\A_1} \geq t \right) \leq \Pr\left(B_x^{\A_2} \geq t \right)\label{eq:dom}
\end{equation}
for $1\leq x \leq n$ and $t \geq 0$.
\end{definition}
We note that domination is a stronger concept than majorization and  that if $\A_1$ is dominated by $\A_2$ then the total number of balls in $\A_1$ at the end of the process may be less than that in $\A_2$.

\begin{definition} 
Let  $1\leq x_0 \leq n$. By $\LA $, we denote an allocation process starting with $n$ empty bins in which each ball chooses a bin i.u.r., say bin $x$ (the $x$th most loaded bin), and  is placed into the bin only if $x > x_0$ and discarded if $x \leq x_0$.
\end{definition}

\section{Key Properties of $(k,d)$-choice}\label{sec:properties}
In this section, we list useful properties of $(k,d)$-choice which will be used later. First,  we need the following lemma.
\begin{lemma}\label{lmm:newDominance} 
Let $\{X_1, \ldots, X_n\}$ and $\{Y_1, \ldots, Y_n\}$ be sets of Bernoulli random variables and
let $\{\omega_r\}_{r=1}^m$ be  a sequence of random variables in the range $\{1, 2, \ldots, n\}$. Suppose that
$X_r= X_r(\omega_1, \ldots, \omega_{r-1}) $ and $Y_r = X_r(\omega_1, \ldots, \omega_{r-1})$, and that
$$\Pr\left(X_t=1 \;|\; ( \omega_1, \ldots, \omega_{t-1}) = \mathbf{u} \right) \leq \Pr\left(Y_t = 1 \;|\; (\omega_1, \ldots, \omega_{t-1}) = \mathbf{u} \right),$$
for any $t \geq 1$ and any $\mathbf{u} \in \{1, 2, \ldots, n\}^{t-1}$.
Then
\begin{align*}
\Pr\left(\sum_{t=1}^n X_t \geq s \right) \leq \Pr \left(\sum_{t=1}^nY_t \geq s\right),
\end{align*} 
for any $s \geq 0$.
\end{lemma}
\subsection{Properties of $(k,d)$-choice}\label{subsec:properties}
\noindent
{\bf Properties of $(k,d)$-choice:} Let $1\leq  k<d \leq n$ and $\alpha \geq 1$. 
\begin{enumerate}[(i)]
\item $A_{\sigma}(k,d) \equiv \A(k,d)$, for any choice of $\sigma$.
\item  $\A(k, d + \alpha)   \leq_{mj} \A(k, d)$.
\item $\A(k-\alpha,d)  \leq_{mj} \A(k, d)$,  if $\alpha < k$.
\item $\A(\alpha k,\alpha d) \leq_{mj} \A(k,d)$.
\item $\A(k,d)  \leq_{mj} \A(k+\alpha, d+\alpha)$.
\end{enumerate}
\begin{proof}
Basic techniques frequently used in this Lemma  are majorization and coupling arguments (See \cite{ABKU94, BCSV00} for background of majorizaton and coupling).
The first four properties are intuitively obvious and can be proved by natural coupling, whereas the last property (v) needs a more sophisticated coupling argument.
We provide a sketch of proof for each of  part (i) - (iv) and detailed analysis for (v).

{\bf Part (i):} 
Consider the following natural coupling for $\A_{\sigma}(k,d)$ and $\A(k, d)$:  In each round $r=1, 2, \ldots, n/k$,  the same set of $d$ random bins are chosen to probe for both $\A_{\sigma}(k,d)$ and $\A(k,d)$. For any permutation $\sigma_{r}$ of $\{1, 2, \ldots, k\}$,  the number of balls in the $x$ most loaded bins for each processor are equal at the end of round $r$ . That is,
$B_{\leq x}^{\A(k,d)}(rk) \leq B_{\leq x}^{\A(k,d)}(r)$ holds under this coupling.

{\bf Part (ii):} 
Assume that, in each round $r$,  a set $S_r$ of $d+\alpha$ random bins and  a random subset of $S_r$ with $d$ bins are chosen to probe for $\A(k,d+\alpha)$ and $\A(k,d)$, respectively.  Under this  coupling, $B_{\leq x}^{\A(k,d+\alpha)}(r) \leq B_{\leq x}^{\A(k,d)}(r)$ holds with certainty.

{\bf Part (iii):} 
The property (ii) is obtained from (v) and (i) as follows:
$$\A(k-\alpha, d) \leq_{mj} \A(k, d+\alpha) \leq_{mj} \A(k,d).$$

{\bf Part (iv):}
We define a coupling that links one round for $\A(\alpha k, \alpha d)$ and  $\alpha$ rounds for $\A(k,d)$ as follows.
Suppose that  a set $S_r$ of $\alpha d$ random bins have been selected in the beginning of round $r$ for $\A(\alpha k, \alpha d)$. Partition the set $S_r$ into $\alpha$ random subsets with equal size, each of which is used  as a set of $d$ random bins  in each of $\alpha$ rounds for  $\A(k, d)$.
One can show that under this coupling
$B_{\leq x}^{\A(\alpha k, \alpha d)}(r) \leq B_{\leq x}^{\A(k, d)}(\alpha r)$.

{\bf Part (v):} 
It suffices to show that 
$$\A_{\sigma}(k, d) \leq_{mj} \A_{\pi}(k+1, d+1),$$
for some $\sigma$ and $\pi$.
For  fixed $1\leq t\leq n$, let $W_{x}^{\A_{\sigma}(k, d) }(t) = B_{\leq x}^{\A_{\sigma}}(k, d) (t) - B_{\leq x}^{\A_{\sigma}(k, d) }(t-1)$.  That is, $W_{x}^{\A_{\sigma}(k, d) }(t)$ is a Bernoulli random variable which is $1$ if and only if the bin containing $t$ is one of the $x$ most loaded bins  at time $t$ (i.e. right after the $t$th ball is placed).
Similarly, let  $W_{x}^{\A_{\pi}(k+1, d+1) }(t) = B_{\leq x}^{\A_{\pi}(k+1, d+1) }(t) - B_{\leq x}^{\A_{\pi}(k+1, d+1) }(t-1)$.
We will show that, for any $1\leq x \leq n$, we have
\begin{align}
\Pr\left (W_{x}^{\A_{\sigma}(k,d)}(t) = 1 \;| \; (\omega_1, \ldots, \omega_{t-1}) = \mathbf{u}\right) \leq 
\Pr\left (W_{x}^{\A_{\pi}(k+1,d+1)}(t) = 1 \;| \; (\omega_1, \ldots, \omega_{t-1}) = \mathbf{u}\right), \label{eq:W}
\end{align}
where $\omega_j$ represents the bin that received  ball $j$ and $\mathbf{u} \in \{1, 2, \ldots, n\}^{k-1}$.
Note that 
$B_{\leq x}^{\mathcal{A_{\sigma}}(k,d)} = \sum_{t=1}^nW_{x}^{\A_{\sigma}(k,d)}(t)$ and $B_{\leq x}^{\mathcal{A_{\pi}}(k+1,d+1)} = \sum_{t=1}^nW_{x}^{\A_{\pi}(k+1,d+1)}(t)$. By  Lemma~\ref{lmm:newDominance}, the inequality (\ref{eq:W}) implies  
\begin{align}
\Pr \left(B_{\leq x}^{\A_{\sigma}(k,d)} \geq s \right) \leq \Pr \left(B_{\leq x}^{\A_{\pi}(k+1, d+1)} \geq s \right). \notag
\end{align}

For the rest of proof,  we specify the processes $\A_{\sigma}(k, d)$ and $\A_{\pi}(k+1, d+1)$
and define a coupling running on $\A_{\sigma}(k, d)$ and $\A_{\pi}(k+1, d+1)$ in which (\ref{eq:W}) holds.  We assume that $n=k(k+1)$ without loss of generality.

Definition of $\A_{\pi}(k+1,d+1)$:
For $y \in \{1, 2, \ldots, k+1\}$, let $H_{y}$ denote the set $\{1, 2, \ldots, k+1\}\setminus \{y\}$.
Define
$$\Omega(k+1, d+1) = \{(H_{y_1}, \ldots, H_{y_{k+1}}) \;|\; (y_1,  \ldots, y_{k+1}) \text{  is a permutation of } \{1, 2, \ldots, k+1\} \}.$$
We view the set $\Omega(k+1,d+1)$ as the space of all possible choices that $k+1$ balls have in each round for $\A_{\pi}(k+1,d+1)$. 
The $\A_{\pi}(k+1,d+1)$ process begins by choosing a permutation $ (y_1, \ldots, y_{k+1})$ of $\{1, \ldots, k+1\}$ randomly and hence determines a list of sets
$(H_{y_1}, \ldots, H_{y_{k+1}}) \in \Omega(k+1, d+1)$. 
In each round $r = 1, 2, \ldots, k$, a set $S_r$ of $d+1$ random bins are selected. Let $\pi_r = (y_{r+1}, y_{r+2}, \ldots, y_{k+1}, y_1, y_2, \ldots, y_r)$ be a permutation of $\{1, 2, \ldots, k+1\}$ by which each of $k+1$ balls in round $r$ is allocated. That is,   the $s$th ball  in the round  is placed into the $y_{r+s}$th least loaded bin if $1\leq s \leq k-r+1$, and  is placed into $y_{s-(k-r+1)}$th least loaded bin if $k-r+2 \leq s \leq  k+1$. Let $\pi = (\pi_1, \ldots, \pi_k)$.

Definition of $\A_{\sigma}(k,d)$:
We view $\A(k,d)$ as a $(d+1)$-random-bins-model, making $\A(k,d)$ comparable to $\A(k+1, d+1)$ as follows.
In each round $r = 1, 2, \ldots, k+1$,  a set of $d+1$ random bins (rather than $d$ bins) is selected first. Then one of the bins is chosen randomly and removed, and then $k$ balls fall into the $k$ least loaded among the remaining $d$ bins. Clearly this process is equivalent to $\A(k,d)$. We describe this procedure formally as follows. 
For $x \in \{1, 2, \ldots, d+1\}$, define $G_x$ to be the set 
\begin{align*}
G_x = \begin{cases}
\{1, \ldots, k+1\} \setminus \{x\}, \quad  &\text{ if } x \leq k+1 \\
\{1, 2, \ldots, k\},  &\text{ if } x > k+1
\end{cases}
\end{align*}
Let
$\Omega(k, d) = \{(G_{x_1}, \ldots, G_{x_{k+1}}) \;|\; (x_1, \ldots, x_{k+1}) \in \{1, 2, \ldots, d+1\}^{k+1} \}$. 
the $\A_{\sigma}(k,d)$ process starts by choosing a vector $(x_1, \ldots, x_{k+1}) \in \{1, 2, \ldots, d+1\}^{k+1} $ randomly to determine a list of sets $(G_{x_1}, \ldots, G_{x_{k+1}}) \in \Omega(k,d)$.
In each round $r=1, 2, \ldots, k+1$,  choose a set $S_r$ of $d+1$ random bins to probe and a permutation $\sigma_r = (i_1, i_2, \ldots, i_{k})$ of $G_{x_r}$.  Each ball $1 \leq s \leq k$
(that belongs to round $r$) is placed sequentially  into the $i_s$th least loaded bin in $S_r$. In the following coupling process, we choose $\sigma_r$ based on $\pi$ used in the definition of  $\A_{\pi}(k+1, d+1)$.

Coupling:
Fix $1\leq t \leq n$ first. We define a coupling for $\A_{\sigma}(k,d)$ and $\A_{\pi}(k+1, d+1)$.
Assume that ball $t$ belongs to round $r$ for $\A_{\sigma}(k,d)$. That is,  $ (r-1)k < t \leq rk$.  
Suppose that  the $\A_{\pi}(k+1, d+1)$ process begins by choosing a permutation $(y_1, \ldots, y_r, \ldots, y_{k+1})$ of $\{1, \ldots, k+1\}$ randomly. Then the corresponding  process of $\A_{\sigma}(k,d)$    chooses a vector $(x_1, \ldots, x_r, \ldots, x_{k+1}) \in \{1, 2, \ldots, d+1\}^{k+1} $ randomly under the following restriction on $x_r$ (and no restrictions on other entries): 
\begin{align*}
x_r = 
\begin{cases}
y_r,  \text{ with probability } \frac{k+1}{d+1}\\
\text{one of $\{k+2, \ldots, d+1\}$ chosen randomly}, \text{with probability }\frac{d-k}{d+1}
\end{cases}
\end{align*}
Therefore, either $G_{x_r} = H_{y_r}$ or $G_{x_r} = \{1, 2, \ldots, k\}$. 
Recall that each of the $k$ balls associated with $H_{y_r}$ under the $\A_{\pi}(k+1, d+1)$ process belongs to either round $r-1$ or round $r$,  and is placed into a bin by the permutation $(y_1,   \ldots, y_{r-1}, y_{r+1}, \ldots,y_{k+1})$ of $H_{y_r}$. To keep the notations simple, we rename the permutation  $(y_1,   \ldots, y_{r-1}, y_{r+1}, \ldots,y_{k+1})$ as $ (j_1, \ldots, j_{k})$. 
The key observation is that, by the choice of $x_r$, there is a permutation  $(i_1, \ldots, i_k)$ of $G_{x_r}$ such that $i_s \leq j_s$ for all $1\leq s \leq k$.  We set $\sigma_r$ to be $(i_1, \ldots, i_k)$ to specify  $\A_{\sigma}(k,d)$.
Let $t_1, \ldots, t_k$ be the balls associated with $H_{x_r}$ and $G_{y_r}$ in both processes. Under $\A_{\pi}(k+1, d+1)$, ball $t_s$, $1 \leq s \leq k$, is placed into $j_s$th least loaded bin (out of the $d+1$ random bins chosen either in round $r-1$ or  $r$). Under $\A_{\sigma}(k,d)$, ball $t_s$ is placed into $i_s$th least loaded bin (out of the $d+1$ random bins chosen in round $r$). Let $t = t_s$ for some $s$. The fact  $i_s \leq j_s $ guarantees the inequality (\ref{eq:W}) as desired.
\end{proof}
\subsection{Proof of Theorem~\ref{thm:2}}
We observe that all the properties of $(k,d)$-choice listed in the previous section hold when the allocation process is extended to the case $m>n$ balls.  Therefore, by properties (iv) and (v), 
$$\A(1, d-k+1) \leq_{mj} \A(k,d) \leq_{mj} \A(1, \lfloor d/k \rfloor)$$
holds regardless the number of balls. Using the result on the heavily loaded case of $d$-choice \cite{BCSV00}, we obtain Theorem~\ref{thm:2}. For $k< d < 2k$, the behavior of the $(k,d)$-choice in the heavily loaded case  remains an open question. The rest of this paper is devoted to prove Theorem~\ref{thm:1}.

\section{Upper Bound Analysis}\label{sec:up}
In this section, we analyze upper bounds on the maximum load $M(k,d,n)$. 
A schematic diagram of the sorted bin load resulting from the $(k,d)$-choice process is shown in Figure~\ref{fig:1}, where $B_x$ represents the number of balls in bin $x$, the $x$th most loaded bin.
We select a suitable constant $\beta_0$ for which we break the maximum bin load into $ B_{\beta_0}$ and  $(B_1 - B_{\beta_0})$, on each of which we derive an upper bound separately. Depending on the range of $k$ and $d_k$  we use different approaches as described in subsequent sections.
\twocolfig{3.7}{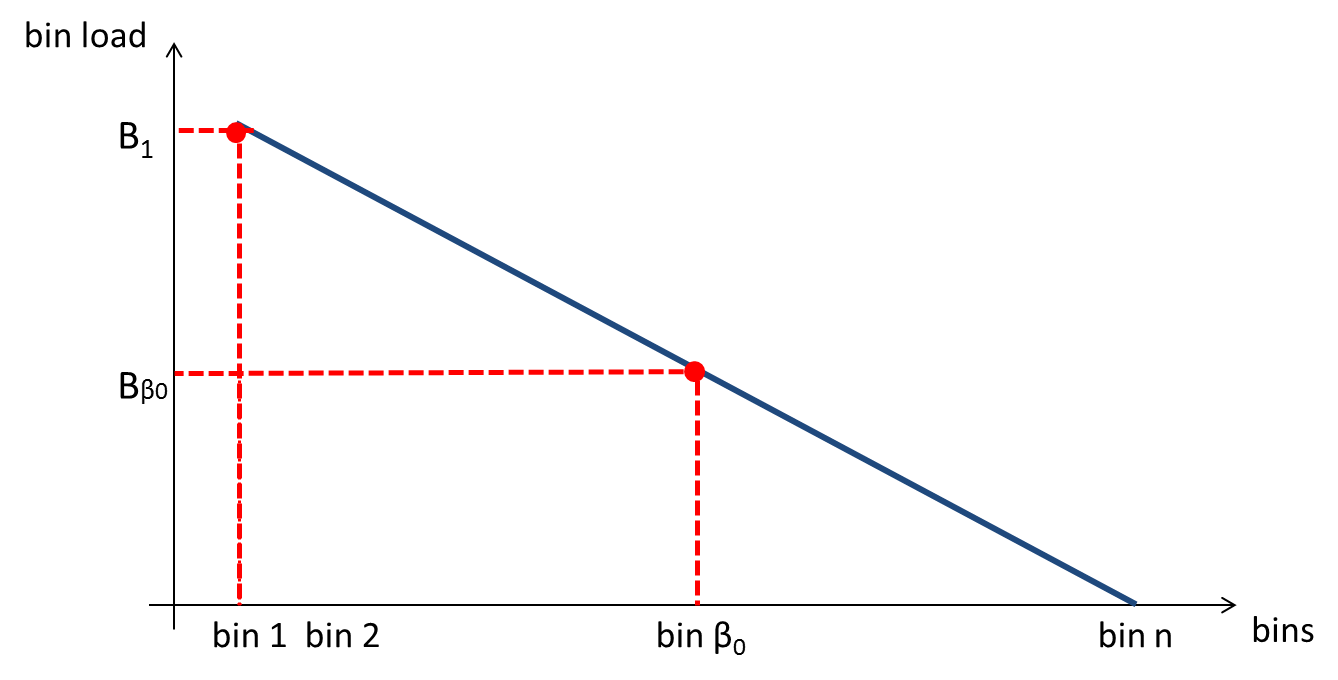}{Sorted bin load vector of $(k,d)$-choice}{fig:1}

\subsection{Upper Bound when $d_k \leq n^{1-\delta}$} \label{sec:up1}
Throughout this subsection, we assume that $d_k \leq n^{1-\delta}$.

\subsubsection{Upper Bound on  $B_{\beta_0}$}  \label{secup1:loadb0}
First, we note  the relation between $\nu_y^{\A}$ and $B_x^{\A}$: $\nu_y^{\A}  \leq x-1  \text{ if and only if  } B_x^{A} \leq y-1.$  Using the following two lemmas and the fact that $\nu_y^{\A} \leq \mu_{y}^{\A}$, we obtain an upper bound on $\nu_y^{A}$ and in turn an upper bound on $B_{x}^{\A}$ for some $x$.
Carefully chosen $x$ will make the bound essentially tight, as proved in Section~\ref{sec:lo1}.

\begin{lemma}\label{lmm:single} For any $y \geq 0$,
\begin{align*}
\Pr\left(\mu_{y}^{\SA} \geq \frac{8n}{y!}\right) &\leq e^{-\frac{n}{12y!}}.
\end{align*}
\end{lemma}
From Lemma~\ref{lmm:single} we can derive an upper bound for $\nu_y^{\SA} \leq \mu_{y}^{\SA}$. 
The following Lemma is a consequence of the property (iii)  listed in Section~\ref{sec:properties}. 
\begin{lemma}\label{lmm:single2}
For any $y \geq 0$ and $t \geq 0$, 
\begin{align*}
\Pr\left(\mu_{y}^{\SA} \geq t \right) &\geq
\Pr\left(\mu_{y}^{\A} \geq t \right).
\end{align*}
\end{lemma}

The main result of this section is formalized as follows.
\begin{theorem}\label{thm:up1}
Let $\beta_0 = \frac{n}{6d_k}$ and $k \leq n^{1-\delta}$.
The number of balls in bin $\beta_0$ resulting from $(k,d)$-choice process is 
\begin{align}
B_{\beta_0}^{\A} \leq \begin{cases} O(1), \qquad &\text{ if } d_k = O(1)\\
(1+o(1))\frac{\ln d_k}{\ln\ln d_k} &\text{ if } d_k \to \infty
\end{cases}
\end{align}
with probability $1-o\left(\frac{1}{n}\right)$.
\end{theorem}
\begin{proof}
By Lemma~\ref{lmm:single},  
$$\mu_{y}^{\SA} \leq \frac{8n}{y!}$$
holds with probability $1-e^{-\frac{n}{12y!}}$. 
Using Lemma~\ref{lmm:single2} and the fact that  $\nu_{y}^\A \leq \mu_{y}^\A$, we obtain
\begin{align}
\nu_{y}^{\A} \leq \frac{8n}{y!}. \label{eq:y1-1}
\end{align}
Let $y_0$ be the smallest $y$ such that 
$\nu_{y}^{\A} \leq \beta_0$. 
Let $y_1 = y_0-1$. By the definition of $y_0$, we have 
\begin{align}
\beta_0 \leq \nu_{y_1}^{\A}. \label{eq:y1-2}
\end{align} 
The inequalities in  (\ref{eq:y1-1}) and (\ref{eq:y1-2}) lead to
$\beta_0 \leq \frac{8n}{y_1!}$, or
$$\frac{n}{6d_k} \leq \frac{8n}{y_1!}.$$
Therefore, we have
\begin{equation}
y_1! \leq 48 d_k. \label{eq:y1-3}
\end{equation}
Using Stirling's formula we solve the above inequality for $y_1$ to obtain
\begin{align*}
y_1 \leq 
(1+\epsilon(d_k))\frac{\ln 48 d_k}{\ln\ln 48 d_k},
\end{align*}
where 
$\epsilon(d_k) = \frac{\ln\ln\ln (48d_k)}{\ln\ln (48d_k)}$.
Note that, since $d_k \leq n^{1-\delta} \ll \frac{n}{\text{polylog }n}$, (\ref{eq:y1-3}) guarantees   
$y_1! \ll n/\ln n$ and hence  $1-e^{-\frac{n}{12y_1!}} \geq 1 - o(1/n)$. Therefore, with probability
$1-o(1/n)$,
\begin{align*}  
y_0 = y_1+1 &\leq (1+\epsilon(d_k))\frac{\ln 48 d_k}{\ln\ln 48 d_k}+1 \\
&\leq \begin{cases}
O(1), \quad &\text{ if } d_k = O(1)\\
(1+o(1))\frac{\ln d_k}{\ln\ln d_k}, &\text{ if } d_k \to \infty.
\end{cases}
\end{align*}
Since  $\nu_{y_0}^{\A} \leq \beta_0$ implies $B_{\beta_0}^{\A}  \leq y_0$, we complete the proof.
\end{proof}

\subsubsection{Upper Bound on $B_1-B_{\beta_0}$} \label{secup1:loaddiff1}
We revisit the layered induction approach presented by the authors in \cite{ABKU94, Power96} (For example, see pages 9 - 13 in \cite{Power96}.) The key formulation in their analysis is the recursive definition for $\nu_{y+1}^{\A(1, d)}$ expressed in terms of $\nu_y^{\A(1, d)}$, where  $\nu_y^{\A(1, d)}$ is the number of bins with load at least $y$. Azar et al. \cite{ABKU94} and Mitzenmacher \cite{Power96}  showed that the sequence of $\{\nu_y^{\A(1, d)}\}_{y \geq 0}$ decreases doubly exponentially (with high probability). Our approach here is similar to the existing analysis. In the $(k,d)$-choice context, however,  the interplay among $k$ balls within a round  in addition to the dependencies between different rounds pose several challenges. For example,  $\nu_{y+1}^{\A(k,d)}$ depends not only on $\nu_y^{\A(k,d)}$ but also on some other bins with load less than $y$; if some bins with less than $y$ balls are sampled multiple times, they may receive multiple balls and in turn become another source that increases the value of $\nu_{y+1}^{\A(k,d)}$. Furthermore, the random variables we deal with take on values in the range $0, 1, \ldots k$, and therefore the Chernoff bounds on the sum of Bernoulli random variables will no longer apply to our case. In addition,  the Chernoff-Hoeffding bounds that hold for random variables in a large range are not strong enough to guarantee the tight bound we desire. 

\begin{lemma}\label{lmm:Xr}
Let $X_{r}(y+1)$ represent the number of balls placed in the $r$th round of $(k,d)$-choice with height at least $y+1$. 
For $1\leq j \leq k$, we have
\begin{align}
\Pr\left(X_r(y+1) \geq j \;|\; \omega_{1}, \ldots, \omega_{r-1}, \nu_y  \right) &\leq {d \choose d-k+j} \left(\frac{\nu_y}{n} \right)^{d-k+j}, \label{eq:Xr}
\end{align}
where $\omega_i$ denotes the bins that received a ball in round $i$ and $\nu_y = \nu_y^{\A}$.
\end{lemma}
In the following two lemmas, we discuss a Chernoff-type tail bound that holds on the sum of  non-Bernoulli random variables under a specific condition.
\begin{lemma}\label{Chernoff}
Let $Y_r$ be independent random variables with $p_j = \Pr \left(Y_r=j \right)$ and $\sum_{j=0}^kp_j = 1$.
If $\{p_j\}_{j=1}^k$ is decreasing by (at least) a factor of $\eta > 1$ \footnote{That is, $p_1 \geq \eta \cdot p_2 \geq \eta^2 p_3 \geq \ldots \geq \eta^{k-1}p_k$.}, then 
the following results hold.  \\
i) For $\delta>0$, we have
\begin{align*}
\Pr \left(\sum_{r=1}^{m}Y_r \geq (1+\delta)p_1m \right) \leq 
\left( \frac{e^{\delta}}{(\eta(1+\delta)/(\eta+1+\delta))^{1+\delta}}\right)^{p_1m}.
\end{align*}
ii) If $\eta = 2e$, then
\begin{align}
\Pr\left(\sum_{r=1}^{m}Y_r  \geq 2e p_1m \right) \leq e^{-p_1m}.\label{2e}
\end{align}
\end{lemma}

\begin{lemma}\label{lmm:tailbound}
Let each of $\{X_r\}_{r=1}^m$  and  $\{Y_r\}_{r=1}^m$ be a sequence of random variables in the range $\{0, 1, \ldots, k\}$. Let $\{\omega_r\}_{r=1}^m$ be  a sequence of random variables in an arbitrary domain. Suppose that
$X_r= X_r(\omega_1, \ldots, \omega_{r-1}) $ and  $Y_r$ are independent. If
\begin{align}
\Pr \left(X_r = j \;|\; \omega_1,  \ldots, \omega_{r-1} \right) \leq \Pr \left(Y_r = j \right) \label{cond}
\end{align}
holds for all $1 \leq j  \leq k$,
then
\begin{align}
\Pr \left(\sum_{r=1}^{m}X_r \geq t\right) \leq \Pr\left(\sum_{r=1}^{m}Y_r \geq t\right), \label{upper-bound}
\end{align}
for any $t \geq 0$.
\end{lemma}
Now we are ready to prove the following result .
\begin{theorem} \label{thm:up2}
Let $\delta = \frac{\ln\ln\ln n}{\ln\ln n}$ and $\beta_0 = \frac{n}{6d_k}$.  If $d_k \leq n^{1-\delta}$, then
the load difference between bin $1$  and bin $\beta_0$ at the end of the $(k,d)$-choice process  is 
$$
B_1^{\A} - B_{\beta_0}^{\A} \leq \frac{\ln\ln n}{\ln (d-k+1)} + O(1)
$$
with probability $1- o\left(\frac{1}{\text{polylog } n}\right)$.
\end{theorem}
\begin{proof}
Fix $y$ and $r$. We call a ball with height at least $y+1$ a {\it high} ball. Let $X_r(y+1)$ denote the number of {\it high} balls placed in round $r$.
Then
$$
\mu_{y+1} = \sum_{r=1}^{n/k}X_r(y+1).
$$ 
We construct a sequence $\{\beta_i\}$ where $0 \leq i \leq i^*$ for some $i^* > 0$ recursively as follows.
\begin{align}
\beta_{0} &= \frac{1}{6}\frac{n}{d_k} \notag \\
\beta_{i+1} &= 6\frac{n}{k}{d \choose d-k+1}\left(\frac{\beta_{i}}{n} \right)^{d-k+1}, \quad i \geq 0. \label{rel}
\end{align}
Recall that,  in Theorem~\ref{thm:up1}, we showed that
$\nu_{y_0} \leq \beta_0$ with probability $1-o(1/n)$, where 
$y_0 = O(1)$ if $d_k = O(1)$ and $y_0 \leq (1+o(1))\frac{\ln d_k}{\ln\ln d_k}$. 
We prove  that the following properties hold with probability $1-o(1)$:
\begin{align}
\nu_{y_0+i} &\leq \beta_i,   \label{eq:up2-1} \\
i^* &\leq \frac{\ln\ln n}{d-k+1}, \label{eq:up2-2}\\
M(k,d,n) &\leq y_0+i^*+2. \label{eq:up2-3} 
\end{align}

The analysis of the theorem is involved and lengthy and therefore we divide the rest of the proof into three parts each of which proves one of the inequalities (\ref{eq:up2-1}) to (\ref{eq:up2-3}).
\\
\\
{\bf Part A:}
We show that, for $0 < i \leq i^*$, $\nu_{y_0+i} \leq \beta_i$ with probability $1-O(\frac{i}{n})$.\\
Fix $i\geq 0$ and $r \leq n/k$.  Recall that $X_{r}(y_0+i+1)$ represents the number of balls placed in the $r$th round with heights at least $y_0+i+1$. Then
$$
\mu_{y_0+i+1} = \sum_{r=1}^{n/k}X_r(y_0+i+1).
$$ 
Let $\mathcal{E}_{i}$ denote the event that $\nu_{y_0+i} \leq \beta_i$.
 and let $\omega_r$ represent the bins that received a ball in the $r$th round. 
By Lemma~\ref{lmm:Xr}, 
 \begin{align*}
\Pr \left(X_{r}(y_0+i+1) \geq j \;|\; \omega_1, \ldots, \omega_{r-1}, \mathcal{E}_i \right)  &\leq  p_j.
\end{align*}
where
$$p_j =  {d \choose d-k+j}\left(\frac{\beta_i}{n}\right)^{d-k+j}.$$
Let $Y_r$ represent an independent random variable such that 
\begin{align*}
\Pr \left(Y_r = j \right) &= p_j, \quad j = 1, 2, \ldots, k\\
\Pr \left(Y_r = 0 \right) &= 1-(p_1+\ldots+p_k).
\end{align*}
Note that $\{p_j\}_{j=1}^k$ is decreasing by at least a factor of $6$;
\begin{align*}
p_{j+1} &= p_j \frac{k-j}{d-k+j+1}\frac{\beta_i}{n} \\
&\leq p_j\frac{d}{d-k}\frac{\beta_0}{n}\\
&\leq p_j\frac{1}{6}.
\end{align*}
We have shown that
\begin{align*}
\Pr \left(X_{r}(y_0+i+1) = j \;|\; \omega_1, \ldots, \omega_{r-1}, \mathcal{E}_i \right)  &\leq  \Pr \left(Y_r=j \right),
\end{align*}
and  $p_j = \Pr(Y_r = j)$ decreases by a factor of at least $2e$.
Applying Lemma~\ref{lmm:tailbound} and  Lemma~\ref{Chernoff},  we have
\begin{align*}
\Pr\left(\nu_{y_0+i+1} > \beta_{i+1} \;|\; \mathcal{E}_i\right) & \leq \Pr\left(\mu_{y_0+i+1} > \beta_{i+1} \;|\; \mathcal{E}_i\right) \\
&=\Pr\left(\sum_{r=1}^{n/k}X_r(y_0+i+1) > 2ep_1n/k     \;|\; \mathcal{E}_i \right ) \\
&\leq \Pr\left(\sum_{r=1}^{n/k}Y_r > 2ep_1n/k \right), \quad \text{ by Lemma~\ref{lmm:tailbound}} \\ 
&\leq e^{-p_1n/k}, \quad \text{ by Lemma~\ref{Chernoff}}.
\end{align*}
Now we choose $i^*$ to be the  largest $i$ such that $\beta_{i} \geq 6\ln n$.
Let $i$ be such that $0 \leq i < i^*$. Since
$\beta_{i+1} = 6p_1n/k  \geq 6\ln n $, we have
$p_1n/k \geq \ln n$. Therefore, 
\begin{align}
\Pr \left(\nu_{y_0+i+1} > \beta_{i+1} \right)&=\Pr\left((\nu_{y_0+i+1} > \beta_{i+1}) \;|\; \mathcal{E}_i\right) \Pr\left(\mathcal{E}_i \right) +\Pr\left(\mathcal{E}_i^{c} \right) \notag\\
&\leq \frac{1}{n} + \Pr \left(\nu_{y_0+i} > \beta_i\right)
 \label{beta-rec}
\end{align}
By applying the formula (\ref{beta-rec}) recursively, we have
\begin{align*}
\Pr \left(\nu_{y_0+i+1} > \beta_{i+1} \right) 
&\leq \frac{1}{n} +\Pr \left(\nu_{y_0+i} > \beta_{i}\right)\\
&\leq \frac{2}{n} + \Pr \left(\nu_{y_0+i-1} > \beta_{i-1}\right)\\
& \leq \ldots\\
& \leq \frac{i+2}{n}.
\end{align*}
{\bf Part B:}  We show that $i^* \leq \frac{\ln\ln n}{\ln (d-k+1)}$ with probability $1-o(1)$.\\
Letting 
$F = \frac{6}{k}{d \choose d-k+1}2^{d-k+1}$,
we rewrite (\ref{rel}) as 
\begin{align}
\frac{\beta_{i+1}}{n} = F(\frac{\beta_i}{n})^{d-k+1}. \label{induction}
\end{align}
Applying induction to the relation (\ref{induction}), we have
\begin{align*}
\frac{\beta_{i}}{n} &= F^{1+(d-k+1)+\ldots+(d-k+1)^{i-1}} \left(\frac{\beta_{0}}{n}\right)^{(d-k+1)^i}\\ 
&= F^{\frac{(d-k+1)^{i}-1}{d-k}} \left(\frac{\beta_{0}}{n}\right)^{(d-k+1)^i}.
\end{align*}
Multiplying $F^{1/(d-k)}$ to each side of the above equation yields
\begin{align}
F^{\frac{1}{d-k}}\frac{\beta_{i}}{n} &= \left(F^{\frac{1}{d-k}}\frac{\beta_{0}}{n}\right)^{(d-k+1)^i}. \label{rec}
\end{align}
Using the fact  that
\begin{align*}
&F = \frac{6}{d-k+1}{d \choose d-k}, \quad \text{and}
\end{align*} 
we bound $F^{1/(d-k)}$ as 
\begin{align}
\frac{d_k}{2} \leq F^{\frac{1}{d-k}}  \leq 3d_k. \label{F}
\end{align}
From (\ref{rec}) and (\ref{F}), we obtain 
\begin{align}
\frac{d_k}{2} \frac{\beta_{i}}{n} &\leq \left(3d_k\frac{\beta_{0}}{n}\right)^{(d-k+1)^i}=\left(\frac{1}{2}\right)^{(d-k+1)^i}. \label{eq:betai*}
\end{align}
Using $\beta_{i^*} \geq 6\ln n$
and  replacing $i$ by $i^*$ in the above inequality, we have
$$ \frac{3 d_k\ln n}{n} \leq \left(\frac{1}{2}\right)^{(d-k+1)^{i^*}}$$
and hence
\begin{align*}
i^* &\leq \log_{d-k+1}\log_2 \left(\frac{n}{ 3d_k  \ln n}\right) \\
&\leq  \frac{\ln\ln n}{\ln (d-k+1)}.
\end{align*}
{\bf Part C}: $\nu_{y_0+i^*+2} =0$ with probability $1-o(1)$\\ 
We have shown that $\nu_{y_0+i^*+1} \leq m$ with probability $1-O\left(\frac{\ln\ln n}{n}\right)$.  
Let $X_r(y_0+i^*+2)$ be the number of balls placed in round $r$ with height at least $y_0+i^*+2$. Using Lemma~\ref{lmm:Xr} again,
\begin{align*}
\Pr \left(X_r(y_0+i^*+2) \geq 1 \;|\; \omega_1, \ldots, \omega_{r-1}, \nu_{y_0+i^*+1}\leq 6 \ln n \right) \leq
p_1,
\end{align*}
where $p_1 = {d \choose d-k+1}\left(\frac{6\ln n}{n}\right)^{d-k+1}$.
By the union bound, 
\begin{align*}
\Pr \left( \nu_{y_0+i^*+2} \geq 1 \;|\; \nu_{y_0+i^*+1} \leq 6 \ln n \right) 
&\leq \Pr \left(\sum_{r=1}^{n/k}X_r(y_0+i^*+2) \geq 1 \;|\; \nu_{y_0+i^*+1} \leq 6\ln n \right)\\
&\leq \frac{\frac{n}{k} p_1 }{\Pr \left( \nu_{y_0+i^*+1} \leq 6\ln n \right)},
\end{align*}
and hence
\begin{align*}
\Pr\left( \nu_{y_0+i^*+2} \geq 1 \right) &\leq \frac{n}{k}p_1 + \Pr \left (\nu_{y_0+i^*+1} > 6\ln n \right). 
\end{align*}
Using the fact that
${d \choose d-k+1} \leq \left(\frac{3d}{d-k+1}\right)^{d-k+1} \leq \left(3d_k\right)^{d-k+1}$,
\begin{align*}
\frac{n}{k} p_1 
&=  \frac{n}{k}{d \choose d-k+1} \left( \frac{6\ln n}{n}\right)^{d-k+1}\\
&\leq \frac{n}{k} \left( \frac{18 d_k \ln n}{n} \right)^{d-k+1} \\
&\leq \left( \frac{18 \ln n}{n^{\delta}}\right)^2 = o\left( \frac{1}{\text{polylog} n}\right)
\end{align*}
We have shown that
\begin{align*}
\Pr\left( \nu_{y_0+i^*+2} \geq 1 \right) &\leq o\left(\frac{1}{\text{polylog } n}\right)
\end{align*}
and hence
the maximum load is at most
$
y_0+i^*+2 
$
with probability $1- o\left(\frac{1}{\text{polylog } n}\right)$.
\end{proof}

From Theorem~\ref{thm:up1} and Theorem~\ref{thm:up2}, we derive an upper bound on the maximum load as follows.
\begin{corollary}
If $d_k \leq n^{1-\delta}$, then
the maximum load of $(k,d)$-choice  is 
\begin{align*}
M(k,d,n) \leq \begin{cases}
\frac{\ln\ln n}{\ln d-k+1} + O(1), \quad &\text{ if } d_k = O(1)\\
\frac{\ln\ln n}{\ln d-k+1} + (1+o(1))\frac{\ln d_k}{\ln\ln d_k}, &\text{ if } d_k \to \infty,
\end{cases}
\end{align*}
with probability $1- o\left(\frac{1}{\text{polylog } n}\right)$.
\end{corollary}

\subsection{Upper Bound  when $d_k \geq n^{1-\delta}$}\label{sec:up2}
It remains to show that the upper bounds stated in Theorem~\ref{thm:1} hold for when $d_k \geq n^{1-\delta}$. 
\begin{theorem}\label{thm:up4}
The upper bounds on $M(k,d,n)$ in Theorem~\ref{thm:1} hold for $d_k \geq n^{1-\delta} $.
\end{theorem}
\begin{proof}
Since  $\frac{\ln\ln n}{\ln d-k+1} = o(1) \frac{\ln d_k}{\ln\ln d_k}$ for $d_k \geq n^{1-\delta} $, it suffices to show that the maximum load is at most $\left(1+o(1)\right)\frac{\ln d_k}{\ln\ln d_k}$. 
By Lemma~\ref{lmm:single2}, the maximum load of $(k,d)$-choice is bounded by the maximum load of single choice w.h.p.  Using the well known result on the maximum load of the classic single choice (see \cite{RS98}), 
$$M(k,d,n) \leq \left(1+2\delta \right)\frac{\ln n}{\ln\ln n}$$
holds with probability $1-o(1)$.
Since $d_k \geq n^{1-\delta},$ 
we have
$$(1+2\delta) \frac{\ln n}{\ln\ln n} \leq (1+4\delta) \frac{\ln d_k}{\ln\ln d_k} , $$ and hence
$$M(k,d,n) \leq (1+4\delta) \frac{\ln d_k}{\ln\ln d_k} = \left(1+o(1)\right)\frac{\ln d_k}{\ln\ln d_k}.$$
\end{proof}

\section{Lower Bound on the Maximum Load}\label{sec:lo}
We provide matching lower bounds on the maximum load of the $(k,d)$-choice process. Recall that,
by the property (v) in Section~\ref{sec:properties}, we have $\A(1, d-k+1) \leq_{mj} \A(k,d)$. Using the fact from \cite{ABKU94} that  $M(1, d-k+1, n) \geq \frac{\ln\ln n}{\ln (d-k+1)} - O(1)$ holds with probability $1-o(1/n)$, 
we obtain
$$M(k, d, n) \geq \frac{\ln\ln n}{\ln (d-k+1)} - O(1).$$
This lower bound is tight if $d_k = O(1)$. Therefore,  we assume that $d_k \to \infty$ in the rest of this section.
\twocolfig{4.3}{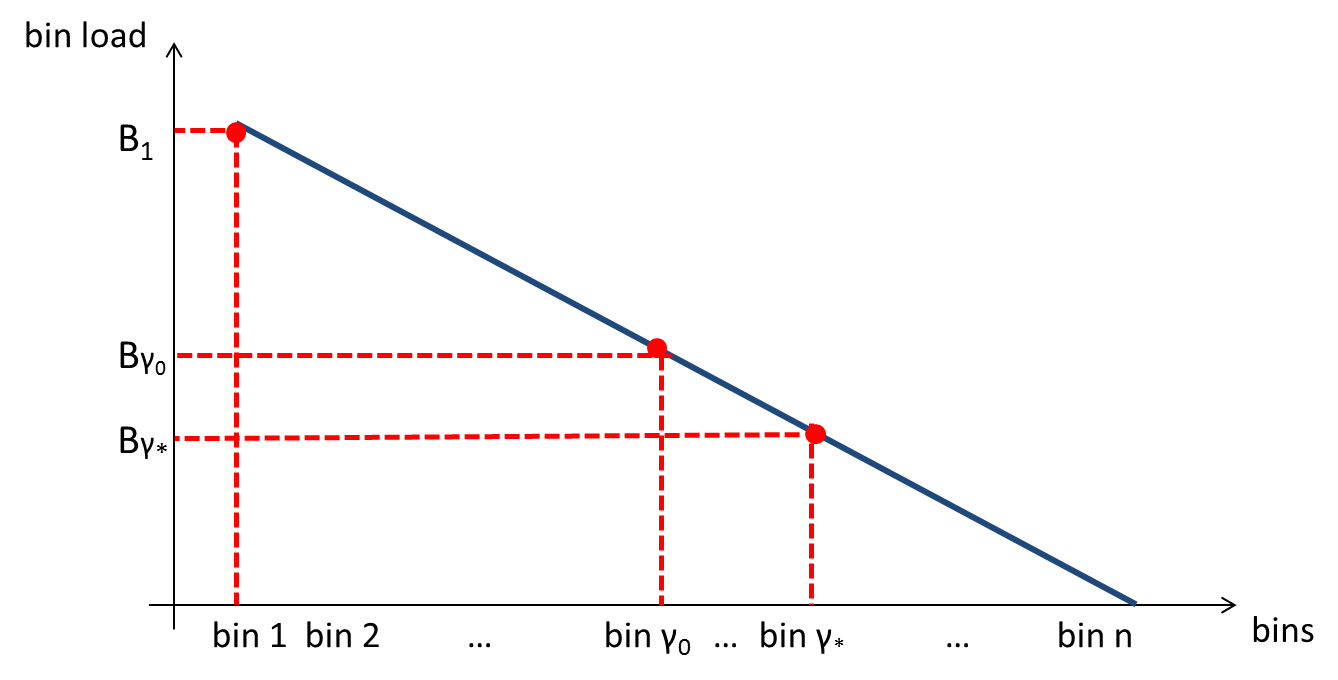}{Sorted bin load vector of $(k,d)$-choice}{fig:2}

As indicated in Figure \ref{fig:2}, the maximum load  of $(k,d)$-choice is at least the sum of bin load $B_{\gamma_*}$ and the load difference $B_{1}-B_{\gamma_0}$, for some $\gamma_* > \gamma_0$ which will be determined later.

\subsection{Lower Bound on $B_{\gamma_*}$} \label{sec:lo1}
In this section, we choose proper $\gamma_*$ and derive a lower bound on 
$B_{\gamma_*}^{\A}$. 
\begin{lemma}\label{lmm:dominated} 
The lower bound analysis is more involved than the upper bound counterpart and needs several new techniques.
Let $\A_1$ and $\A_2$ be serial processes with $n$ empty bins initially. If $p_{x}^{\A_1}(t) \leq p_{x}^{\A_2}(t)$ \footnote{ Recall that $p_{x}^{\A_1}(t)$ denotes the probability that the $t$th ball is placed into bin $x$ resulting from a serialized algorithm $A_1$.} for  $x, t = 1, 2, ..., n$, then $\A_1 \leq_{dm} \A_2$. 
\end{lemma}

\begin{lemma}\label{pr:la}  The $\LA$ process satisfies the following properties.
\begin{enumerate}[(i)]
\item $p_x^{\LA}(t) = \frac{1}{n}$ if $x \geq x_0$, and $p_x^{\LA} = 0$ if $x < x_0$.
\item Either $B^{\LA}_1 = B^{\LA}_{x_0}$, or $B^{\LA}_1 = B^{\LA}_{x_0} + 1$
\item $\LA \leq_{dm} \SA$
\end{enumerate}
\end{lemma}

\begin{lemma}\label{lmm:x1}
For any  $1\leq x_0 < x_1 \leq n$. Then
\begin{align}
\Pr \left(B_{x_1}^{\SA} \geq s \right) \leq \Pr\left(B_{x_0}^{\LA} \geq \left(1-\frac{x_0}{x_1}\right)s-1 \right) \label{eq:x1}
\end{align}
for any $s \geq 0$.
\end{lemma}

\begin{lemma}\label{lmm:gamma}
Let $\gamma_* = 4\frac{n}{d_k}$ and $d_k \to \infty$.  
Then there is a serialization $\A_{\sigma} = \A_{\sigma}(k,d)$ of $\A(k,d)$ such that for each ball $1\leq t \leq n$ and  $x \geq \gamma_*$
$$ p_{x}^{\A_{\sigma}}(t) \geq \frac{1}{n}.$$
\end{lemma}

\begin{corollary}\label{cor:prec} 
Let $\gamma_* =  4\frac{n}{d_k}$.\\
(i) 
\begin{align*}
\mathcal{\SA}_{\gamma_*} &\leq_{dm} \A.
\end{align*}
(ii) 
For any  $\gamma > \gamma_*$. Then
\begin{align*}
\Pr \left(B_{\gamma}^{\SA} \geq s \right) \leq \Pr\left(B_{\gamma_*}^{\A} \geq \left(1-\frac{\gamma_*}{\gamma}\right)s-1 \right), 
\end{align*}
for any $s \geq 0$.
\end{corollary}
\begin{proof}
Part (i) is a direct consequence of Lemma~\ref{lmm:dominated} and Lemma~\ref{lmm:gamma}. Part (ii) is obtained from the result of Part (i) and Lemma~\ref{lmm:x1}.
\end{proof}

The following lemma is used to derive a lower bound on $\nu_y^{\SA}$.
\begin{lemma}\label{lmm:singlelo}
\begin{align*}
\Pr\left(\nu_{y}^{\SA}  \leq  \frac{1}{8}\frac{n}{y!} \right) &\leq e^{-\frac{1}{32}\frac{n}{y!}}.
\end{align*}
\end{lemma}

Now we are ready to state the main result of this section.
\begin{theorem}\label{thm:lo1}
If $d_k \to \infty$ and $\gamma_* =  4\frac{n}{d_k}$, then the load of bin $\gamma_*$ at the end of $(k,d)$-choice process is 
$$B_{\gamma_*}^{\A} \geq (1-o(1))\frac{\ln d_k}{\ln\ln d_k},$$
with probability $1-O(1/n^2)$.
\end{theorem}
\begin{proof}
Let $\gamma = (\ln d_k)\frac{n}{d_k}$.
Let $y_1$ be the smallest $y$ satisfying
$\nu_{y}^{\SA} \leq \gamma$.
Then $B_{\gamma}^{\SA} \geq y_1-1$.
By Lemma~\ref{lmm:singlelo}, we have, with probability $1- \exp(-\frac{n}{32\cdot y_1!})$,
$$\frac{n}{8y_1!} \leq \nu_{y_1}^{\SA}. $$
And hence
$$
\frac{n}{8y_1!} \leq \ln d_k\frac{ n}{d_k}.
$$
Using Stirling's approximation
we obtain
$$y_1 \geq \frac{\ln d_k}{\ln\ln d_k}-O(1),$$ with probability $1- e^{-\frac{1}{32}\frac{n}{y_1!}} \geq 1-O(1/n^2)$.
We have shown that
$B_{\gamma}^{\SA} \geq \frac{\ln d_k}{\ln\ln d_k}-O(1)$ with probability $1-O(1/n^2)$.
Let $y_* = \left(1-\frac{\gamma_*}{\gamma}\right)(y_1-1)-1 \geq (1-o(1))\frac{\ln d_k}{\ln\ln d_k}$. Then
by Corollary~\ref{cor:prec} Part (ii), 
 $B_{\gamma_*}^{\A} \geq y_*$  with probability $1-O(1/n^2)$.
\end{proof}

\subsection{Lower Bound on  $B_1 - B_{\gamma_0}$}\label{sec:lo2}
By finding a lower bound on  the difference $B_1 - B_{\gamma_0}$ for some $\gamma_0$, we complete Theorem~\ref{thm:1}. If $d_k \geq  e^{(\ln\ln n)^3}$, then the lower bound on $B_{\gamma_*}$ presented in Theorem~\ref{thm:lo1} is an asymptotically tight lower bound on the maximum load $B_1$. Therefore, we assume that $d_k \leq e^{(\ln\ln n)^3}$ throughout this section. 

The basic approach used in this section is similar in spirit to the lower bound technique presented in \cite{ABKU94}. 
The following lemma is the Chernoff bound for binomially distributed random variable $B(m,p)$.
\begin{lemma} \cite{ABKU94} \label{lmm:lo}
$$\Pr \left(B(m,p) \leq mp/2 \right) \leq e^{-mp/8}.$$
\end{lemma}
\begin{lemma}\label{lmm:lo1}  \cite{ABKU94}
Let $\omega_1, \omega_2, \ldots, \omega_m$ be a sequence of random variables in an arbitrary domain, and let $Z_1, Z_2, \ldots, Z_m$ be a sequence of Bernoulli random variables with the property that  $Z_r = Z_r(\omega_1, \ldots, \omega_{r-1})$. If
$$
\Pr \left(Z_r=1 \;|\; \omega_1, \ldots, \omega_{r-1} \right) \geq p,
$$ 
then
$$
\Pr\left(\sum_{r=1}^mZ_r > t\right) \geq \Pr \left(B(m, p) > t \right).
$$
\end{lemma}

\begin{theorem} \label{thm:lo2}
Suppose that $d_k \leq e^{(\ln\ln n)^3}$. Let $\gamma_0 = \frac{n}{d}$. The load difference between bin $1$ and bin $\gamma_0$ at the end of the $(k,d)$-choice process is 
\begin{align}
B_{1}^{\A} - B_{\gamma_0}^{\A} \geq \frac{\ln \ln n}{\ln (d-k+1)} - O(1),
\end{align}
with probability $1-o\left(\frac{1}{n}\right)$.
\end{theorem}
\begin{proof}
First, we divide the $n/k$ rounds in the entire course of $(k,d)$-choice  into the following subsets of rounds:
$$\{1, 2, \ldots,R_0\}, \{ R_0+1, R_0+2, \ldots,  R_1\},  \{R_1+1, R_1+2, \ldots, R_2\} , \ldots  $$
where $R_i = \frac{n}{k}\left(1-\frac{1}{2^{i+1}}\right)$.
Recall that $\nu_y(R_i)$ represents the number of bins with at least $y$ balls at the end of the round $R_i$.
We choose $y_0$ be the largest $y$ such that $\nu_y(R_0) \geq \frac{n}{d}$.
Define a sequence $\{\gamma_i\}$ as
\begin{align} 
\gamma_{0} &=  \frac{n}{d},\\
\gamma_{i+1} &= \frac{1}{2^{i+6}}\frac{n}{k}{d \choose d-k+1} \left(\frac{\gamma_i}{n}\right)^{d-k+1},  \quad i \geq 0. \label{eq:gamma}
\end{align}
Let $\mathcal{F}_i$ denote the event that $\nu_{y_0+i}(R_i) \geq \gamma_i$.
We will show that if $\mathcal{F}_i$ holds then $\mathcal{F}_{i+1}$ holds with probability $1-o(1/n^2)$.
\\
{\bf Part A:}
Fix $i \geq 0$. We suppose that $\mathcal{F}_i$ holds.  Let $I_i =\{R_i+1, R_i+2, \ldots, R_{i+1}\}$. For $r \in I_i$,  define Bernoulli random variable $Z_r$ as
\begin{align*}
Z_r = 1 \text{ if and only if } 
	\begin{cases}
	\text{there is a ball in round $r$ with height exactly $y_0+i+1$, or} \\
	\nu_{y_0+i+1}(r-1) \geq \gamma_{i+1}.
	\end{cases}
\end{align*}
Since the total number of balls with height $y_0+i+1$ placed in any round in $I_i$ is no more than $\nu_{y_0+i+1}(R_{i+1})$, by the definition of $Z_r$, we have either $\nu_{y_0+i+1}(R_{i+1}) \geq \sum_{r \in I_i} Z_r$ or $\nu_{y_0+i+1}(R_{i+1}) \geq \gamma_{i+1}$ or both. 
First, we derive a proper lower bound on  
the probability that $Z_r = 1$ given $\mathcal{F}_i$.
For this purpose, we assume that $\nu_{y_0+i}(r-1) \leq (d-k+1)\frac{n}{d}$ without loss of generality 
\footnote{Since $\Pr\left(Z_r = 1  \;|\; \gamma_i \leq  \nu_{y_0+i}(r-1) \leq (d-k+1)\frac{n}{d}\right) \leq  \Pr\left(Z_r = 1  \;|\; \nu_{y_0+i}(r-1) \geq  (d-k+1)\frac{n}{d}\right)$, it suffices to find a lower bound on the probability assuming that $\nu_{y_0+i}(r-1) \leq (d-k+1)\frac{n}{d}$.}
and that $\nu_{y_0+i+1}(r-1) \leq \gamma_{i+1}$ (otherwise, $Z_r = 1$ automatically). 
The probability that there is a ball in the $r$th round with height exactly $y_0+i+1$ is at least
\begin{align}
{d \choose d-k+1} \left(\left( \frac{\nu_{y_0+i}(r-1)}{n}\right)^{d-k+1}-\left(\frac{\nu_{y_0+i+1}(r-1)}{n}\right)^{d-k+1}\right)\left(1-\frac{\nu_{y_0+i}(r-1)}{n}\right)^{k-1}. \label{ineq}
\end{align}
We use the fact that $f(x) = x^{d-k+1}(1-x)^{k-1}$ is increasing on the interval $(0, (d-k+1)/d)$ and that $\frac{\gamma_i}{n} \leq \frac{\nu_{y_0+i}(r-1)}{n} \leq \frac{\nu_{y_0}}{n} \leq \frac{d-k+1}{d}$ to derive 
\begin{align}
\left(\frac{\nu_{y_0+i}(r-1)}{n}\right)^{d-k+1}\left(1-\frac{\nu_{y_0+i}(r-1)}{n}\right)^{k-1} \geq \left(\frac{\gamma_i}{n}\right)^{d-k+1}\left(1-\frac{\gamma_i}{n}\right)^{k-1}. \label{eq:Zr1}
\end{align}
Since $\nu_{y_0+i+1}(r-1) \leq \gamma_{i+1}$, we have
\begin{align}
\left(\frac{\nu_{y_0+i+1}(r-1)}{n}\right)^{d-k+1}\left(1-\frac{\nu_{y_0+i}(r-1)}{n}\right)^{k-1} \leq \left(\frac{\gamma_{i+1}}{n}\right)^{d-k+1}\left(1-\frac{\gamma_i}{n}\right)^{k-1}. \label{eq:Zr2}
\end{align}
From (\ref{eq:Zr1}) and (\ref{eq:Zr2}), we obtain
\begin{align*}
\Pr \left(Z_r = 1 \;|\; \omega_1, \ldots, \omega_{r-1}, \mathcal{F}_i \right) &\geq {d \choose d-k+1} \left( \left(\frac{\gamma_i}{n}\right)^{d-k+1}-  \left(\frac{\gamma_{i+1}}{n}\right)^{d-k+1}  \right) \left(1-\frac{\gamma_i}{n}\right)^{k-1}.
\end{align*}
where $\omega_j$ represent the bins which received balls in the $j$th round.
Using the facts that $\gamma_{i+1} \leq \gamma_i/2$ and  $\left(1-\frac{\gamma_{i}}{n}\right)^{k-1} \geq \left(1-\frac{1}{d}\right)^{k-1} \geq \frac{1}{4}$, we further bound 
\begin{align*}
\Pr \left(Z_r = 1 \;|\; \omega_1, \ldots, \omega_{r-1}, \mathcal{F}_i \right) &\geq \frac{1}{2^3} {d \choose d-k+1} \left(\frac{\gamma_i}{n}\right)^{d-k+1}.
\end{align*}
{\bf Part B:}
Now, we are ready to show that the following properties hold with high probability;
\begin{enumerate} [(1)]
\item $\sum_{r \in I_i}Z_r \geq \gamma_{i+1}$, which implies that $\mathcal{F}_{i+1}$ holds.
\item There is $y^* \geq \frac{\ln \ln n/d_k}{\ln (d-k+1)} - O(1)$ such that
$\nu_{y_0+y^*} (R_{y^*}) \geq  9\ln n$.
\end{enumerate}
Let $p:=\frac{1}{2^3} {d \choose d-k+1}$ and $n_k= \frac{n}{k}$.
Since $\gamma_{i+1} = \frac{1}{2} \frac{n_k}{2^{i+2}}p$, 
applying Lemma~\ref{lmm:lo} and Lemma~\ref{lmm:lo1} yields
\begin{align*}
\Pr \left(\sum_{r \in I_i} Z_r \leq \gamma_{i+1} \;|\; \mathcal{F}_i \right) &\leq \Pr\left(B\left(\frac{n_k}{2^{i+2}}, p\right) \leq \gamma_{i+1}\right)\\
&\leq e^{-n_kp/(8\cdot 2^{i+2})}
\end{align*}
If $n_kp/(8\cdot 2^{i+2}) > 2\ln n$, then we have
\begin{align*}
\Pr \left(\mathcal{F}_{i+1}^c \;|\; \mathcal{F}_i \right) \leq \Pr \left(\sum_{r \in I_i} Z_r \leq \gamma_{i+1}
\;|\; \mathcal{F}_i
\right ) \leq o(1/n^2).
\end{align*}
We find a range of $i$ values such that $e^{-n_kp/(8\cdot 2^{i+2})} = o(1/n^2)$. We solve the following inequality for $i$:
\begin{align*}
\frac{n_kp}{2^{i+2}} &\geq 18 \ln n,
\end{align*}
or equivalently
\begin{align}
\gamma_{i+1} &\geq 9 \ln n. \label{eq:ivalue}
\end{align}
We seek the largest $i$ value for which (\ref{eq:ivalue}) holds.
We apply induction to the recursive definition of $\gamma_i$ given in (\ref{eq:gamma}) and obtain the relation between $\gamma_i$ and $\gamma_0$;
\begin{align*}
\frac{\gamma_{i}}{n}  &= 2^{-\{(i+5) + (i+4)(d-k+1) + (i+3)(d-k+1)^2+\ldots + 6(d-k+1)^{i-1} \}}\\
&\cdot
\left(\frac{1}{k}{d \choose d-k+1} \right)^{\{1+(d-k+1)+(d-k+1)^2+\ldots + (d-k+1)^{i-1}\}} \left(\frac{\gamma_{0}}{n}\right)^{(d-k+1)^i}.
\end{align*}
Using a summation formula, we  have
\begin{align*}
&(i+5) + (i+4)(d-k+1) + (i+3)(d-k+1)^2+\ldots + 6(d-k+1)^{i-1} \\
&= \frac{-i-6+((d-k+1)^i-1)/(d-k)+6(d-k+1)^i}{d-k}\\
&\leq 7(d-k+1)^i.
\end{align*}
Then, we obtain
\begin{align*}
\frac{\gamma_{i}}{n}  &\geq 2^{-7(d-k+1)^i}\left(\frac{1}{k}{d \choose d-k+1}\right)^{\frac{(d-k+1)^i-1}{d-k}} \left(\frac{\gamma_{y_0}}{n}\right)^{(d-k+1)^i}.
\end{align*}
We further bound $\left(\frac{1}{k}{d \choose d-k+1}\right)^{\frac{(d-k+1)^i-1}{d-k}}$ as
\begin{align*}
\left(\frac{1}{k}{d \choose d-k+1}\right)^{\frac{(d-k+1)^i-1}{d-k}}  &= \left(\frac{1}{d-k+1}{d \choose d-k}\right)^{\frac{(d-k+1)^i-1}{d-k}}\\
&\geq  \left(\frac{1}{d-k+1} \left(\frac{d}{d-k}\right)^{d-k} \right)^{\frac{(d-k+1)^i-1}{d-k}}\\
&\geq  \left(\frac{1}{2} \frac{d}{d-k} \right)^{(d-k+1)^i-1}\\
&\geq  \left(\frac{1}{2} \frac{d}{d-k} \right)^{(d-k+1)^i}\frac{1}{d_k}.
\end{align*}
Therefore, we have
\begin{align*}
\frac{\gamma_{i}}{n}  &\geq \left(\frac{1}{2^8}\frac{d}{d-k}\frac{\gamma_{y_0}}{n}\right)^{(d-k+1)^i}\frac{1}{d_k}\\
&\geq \left(\frac{1}{2^8}\frac{1}{(d-k)}\right)^{(d-k+1)^i}\frac{1}{d_k}.
\end{align*}
Now, we solve the inequality for $i$
$$
n\left(\frac{1}{2^8}\frac{1}{(d-k)}\right)^{(d-k+1)^i}\frac{1}{d_k} \geq 9  \ln  n.
$$
to obtain
\begin{align}
i \leq \log_{d-k+1} \log_{2^8(d-k)} \left(\frac{n}{9 d_k \ln n}\right). \label{eq:largesti}
\end{align}
We let $y^*$ be the largest $i$ that satisfies (\ref{eq:largesti}). 
Then 
$\nu_{y_0 + y^*} \geq \gamma_{y^*} \geq 9 \ln n$ holds with probability $1-o(1/n)$ and 
$$ y^*  \geq \log_{d-k+1} \log_{2^8(d-k)} \left(\frac{n}{9 d_k \ln n}\right)-1 = \frac{\ln \ln (n/d_k)}{\ln d-k+1} - O(1).$$
Since $d_k \leq e^{(\ln \ln n)^3}$, we have $y^* \geq  \frac{\ln \ln n }{\ln d-k+1} - O(1)$.
\end{proof}

\section{Proof of Lemmas} \label{sec:lemmas}
In this section, we present lemmas used in previous sections. The proof of Lemma~\ref{lmm:newDominance}  is similar to that of Lemma~\ref{lmm:tailbound} and skipped.

\subsection{Proof of Lemma~\ref{lmm:single} and Lemma~\ref{lmm:singlelo}}
Fix $y \geq 1$. Define random variable $B_{i, y} $ to represent the number of balls in the $i$th bin  whose height at least $y$ after $n$  balls are placed into $n$ bins following the classical single choice balls into bins algorithm. In this proof, we do not assume that bins are sorted by bin load and hence the $i$th bin is not necessarily the $i$th most loaded bin. 
 That is,  
$$B_{i, y} = [B_i-y+1]^{+}$$ 
where $B_i$ is the number of  balls in the $i$th bin and $[z]^{+} = \max\{z, 0\}$. 
Let $L_{i,y}$ be an indicator random variable defined as $L_{i, y} = 1$ if and only if  $i$th bin contains at least $y$ balls.
Then
\begin{align*}
\mu_y^{\SA} &= \sum_{i=1}^n B_{i, y}\\
\nu_y^{\SA} &= \sum_{i=1}^n L_{i, y}.
\end{align*}
For any $y' \geq y \geq 1$, we have
\begin{align*}
\Pr\left(B_{i, y} = y'-y+1 \right) &= {n \choose y'}\left(\frac{1}{n}\right)^{y'}\left(1-\frac{1}{n}\right)^{n-y'}\\
&= {n \choose y'+1}\left(\frac{1}{n}\right)^{y'+1}\left(1-\frac{1}{n}\right)^{n-y'-1} \frac{(n-1)(y'+1)}{n-y'}\\
&\geq {n \choose y'+1}\left(\frac{1}{n}\right)^{y'+1}\left(1-\frac{1}{n}\right)^{n-y'-1} (y'+1)\\
&= (y'+1)\Pr \left(B_{i,y} = y'-y+2 \right)\\
&\geq 2\Pr \left(B_{i,y} = y'-y+2\right).
\end{align*}
Thus
\begin{align}
\Pr \left(B_{i,y} = 1 \right) \leq
\E\left[B_{i,y}\right] &= \sum_{j \geq 1} j\Pr \left(B_{i,y} = j \right) \notag \\
		&\leq \Pr\left(B_{i,y} = 1 \right)\sum_{j \geq 1} j\left(\frac{1}{2}\right)^{j-1} \notag \\
		&\leq4\Pr \left(B_{i,y} = 1 \right). \label{eq:exp}
\end{align}
We can bound $\Pr \left(B_{i, y} = 1 \right)$ as
\begin{align}
\frac{1}{3} {n \choose y}\left(\frac{1}{n}\right)^{y} \leq \Pr\left(B_{i,y} = 1\right) = {n \choose y}\left(\frac{1}{n}\right)^{y}\left(1-\frac{1}{n}\right)^{n-y} \leq  {n \choose y}\left(\frac{1}{n}\right)^{y} . \label{eq:prob}
\end{align}
Combining (\ref{eq:exp}) and (\ref{eq:prob}) we obtain
\begin{align}
\frac{n}{3}{n \choose y}\left(\frac{1}{n}\right)^{y} \leq \E \left[\mu_y^{\SA}\right]    \leq 4n {n \choose y}\left(\frac{1}{n}\right)^{y} \leq \frac{4n}{y!}\label{eq:mu-up}
\end{align}
Since
$
\E\left[L_{i,y}\right] \geq \Pr\left(B_{i,y}=1\right),
$
we have
\begin{align*}
 \E\left[\nu_y^{\SA} \right] &\geq   \frac{n}{3} {n \choose y}(\frac{1}{n})^y \geq 
 \frac{n}{3} \left(\frac{n-y}{n}\right)^y \frac{1}{y!}\\
&  \geq \frac{n}{4y!}, \quad (\text{ if } y \ll \sqrt{n}).
\end{align*}

Next, we show that $\mu_y^{\SA}$ is the sum of Bernoulli random variables which are negatively associated and hence the Chernoff bounds apply to it.
Let $B_{i,j}$ is a $0-1$ random variable which set to $1$ if ball $j$ is placed into bin $i$. Define $B_{i,j,y} = B_{i,j}$ if the height of ball $j$ is at least $y$ and $0$ otherwise.
Then we have $\mu_y^{\SA} =\sum_{i=1}^n\sum_{j=1}^tB_{i,j,y}$. 
Using the fact that $\{B_{i,j}\;|\; i, j \}$ is negatively associated (\cite{Dubhashi96}) and $B_{i,j,y}$ is an increasing function of $B_{i,j}$, the random variables $B_{i,j, y}$ are also negatively associated. 
Similarly,  $\nu_y^{\SA} = \sum_{i=1}^n L_{i,y}$, where $L_{i,y}$ is an indicator function of the event $B_i(t) \geq y$ and hence is an increasing function of $B_i$ which is negatively associated.   
The desired results in Lemma~\ref{lmm:single} and ~\ref{lmm:singlelo} are obtained by applying Chernoff bounds to $\mu_y^{\SA}$ and $\nu_y^{\SA}$.

\subsection{Proof of Lemma~\ref{lmm:Xr}}
Suppose that $d$ random bins have been selected in the beginning of round $r$. If we place $d$ balls into those $d$ bins selected and then take out $d-k$ balls (from those $d$ balls just placed ) with maximum heights, this allocation scheme is equivalent to $(k,d)$-choice. Let $H$ be the set of bins having at least $y$ balls at the end of $(k,d)$-choice process. Then $|H| = \nu_y$.
 In order to make the event $X_r(y+1) \geq j$ occur, at least $d-k+j$ balls must have landed in some bins in $H$. Therefore, by the union bound, 
 $$\Pr\left(X_r(y+1) \geq j \; | \; \nu_y \right) \leq  { d \choose d-k+j } \left(\frac{\nu_y}{n}\right)^{d-k+j}.$$
\subsection{Proof of Lemma~\ref{Chernoff}}
We start with the moment generating function for each $Y_r$:
\begin{align*}
M_{Y_r}(t) &= \E[e^{tY_r}] \\
&= p_0 + p_1e^t+p_2e^{2t} + \ldots + p_ke^{kt}\\
&\leq 1-p_1 + p_1e^t+p_2e^{2t} + \ldots + p_ke^{kt}.
\end{align*}
For $0<t < \ln\eta$, using the fact that $\{p_j\}$ is decreasing by a factor of at least $\eta$,   we have
$p_2e^{2t} + \ldots +p_ke^{kt} \leq p_1e^t(e^t/\eta + (e^t/\eta)^2+\ldots (e^t/\eta)^{k-1}) \leq cp_1e^t$, 
where $c = e^t/(\eta - e^t)$.
Therefore,
\begin{align*}
\E[e^{tY_r}]  &\leq 1-p_1  + (1+c)p_1e^t\\
&\leq 1 + p_1((1+c)e^t-1)\\
&\leq e^{p_1((1+c)e^t-1)}.
\end{align*}
The moment generating function for $Y = \sum_{r=1}^{m}Y_r$ is bounded as 
\begin{align*}
M_{Y}(t) &= \prod_{r=1}^{m} M_{Y_r}(t) 
= \prod_{r=1}^{m}e^{p_1((1+c)e^t-1)}= e^{p_1((1+c)e^t-1)m}.
\end{align*}
Applying Markov's inequality, we have
\begin{align*}
\Pr\left(Y\geq (1+\delta)p_1m \right) &= \Pr \left(e^{tY} \geq e^{t(1+\delta)p_1m}\right)\\
&\leq \frac{\E[e^{tY}]}{e^{t(1+\delta)p_1m}}\\
&\leq \frac{e^{((1+c)e^t-1)p_1m}}{e^{t(1+\delta)p_1m}}
\end{align*}
Set $t = \ln(\eta(1+\delta)/(\eta+1+\delta))$ to be the solution of the equation
$$
(1+c)e^t-1=\delta
$$
to obtain
\begin{align*}
\Pr\left(Y\geq (1+\delta)p_1m\right) \leq 
\left( \frac{e^{\delta}}{(\eta(1+\delta)/(\eta+1+\delta))^{1+\delta}}\right)^{p_1m}.
\end{align*}
In particular, if $\eta >e$, 
choose $\delta = ((e-1)\eta+e) /(\eta-e)$ to satisfy the equation
$$\frac{\eta(1+\delta)}{\eta+1+\delta} = e,$$ and hence
we obtain
\begin{align*}
\Pr\left(Y \geq \frac{e\eta}{\eta-e}p_1m\right) \leq e^{-p_1m}. 
\end{align*}
Finally, choose $\eta = 2e$ in the above inequality to get the result (\ref{2e}).

\subsection{Proof of Lemma~\ref{lmm:tailbound}}

If $m=1$, it is obvious that (\ref{upper-bound}) holds.
Assuming that (\ref{upper-bound}) holds when $m=n-1$,  $n \geq 2$, we prove it holds for $m=n$ too.
Since
\begin{align*}
\Pr\left(\sum_{r=1}^{n}X_r \geq t\right) &= \Pr\left(\sum_{r=2}^{n}X_r  \geq t \;|\; X_1 = 0\right)\Pr \left(X_1 = 0\right)\\
&+  \sum_{j=1}^k\Pr\left(\sum_{r=2}^{n}X_r  \geq t-j \;|\; X_1 = j\right)\Pr \left(X_1 = j \right),
\end{align*}
by induction hypothesis
\begin{align}
\Pr\left(\sum_{r=1}^{n}X_r \geq t\right) &\leq  \Pr\left(\sum_{r=2}^{n}Y_r  \geq t\right)\Pr \left(X_1 = 0 \right)
+  \sum_{j=1}^k\Pr\left(\sum_{r=2}^{n}Y_r  \geq t-j \right)\Pr \left(X_1 = j \right) \label{eq:sum}
\end{align}
Next, we split $\Pr \left(X_1=0 \right)$ into $k+1$ sums as 
\begin{align*}
\Pr\left(X_1 = 0\right) &=\Pr\left(Y_1 = 0 \right) + \Pr\left(X_1=0\right)-\Pr\left(Y_1 = 0\right)\\
&= \Pr \left(Y_1 = 0 \right) + \sum_{j=1}^k\left(\Pr\left(Y_1=j\right)-\Pr\left(X_1 = j\right) \right).
\end{align*}
and rewrite the first term in (\ref{eq:sum})
\begin{align*}
&\Pr\left(\sum_{r=2}^{n}Y_r  \geq t \right)\Pr \left(X_1 = 0 \right) \\
&=  \Pr\left(\sum_{r=2}^{n}Y_r  \geq t \right)\Pr \left(Y_1 = 0 \right) + \sum_{j=1}^k  \Pr\left(\sum_{r=2}^{n}Y_r  \geq t\right )\left(\Pr\left(Y_1=j \right)-\Pr\left(X_1 = j\right) \right)\\
&\leq \Pr\left(\sum_{r=2}^{n}Y_r  \geq t \right)\Pr\left(Y_1 = 0\right) + \sum_{j=1}^k  \Pr\left(\sum_{r=2}^{n}Y_r  \geq t-j \right)\left(\Pr \left(Y_1=j \right)-\Pr\left(X_1 = j\right)\right).
\end{align*}
Therefore, 
\begin{align*}
\Pr\left(\sum_{r=1}^{n}X_r \geq t\right) &\leq  \Pr\left(\sum_{r=2}^{n}Y_r  \geq t \right)\Pr \left(Y_1 = 0 \right) + \sum_{j=1}^k  \Pr\left(\sum_{r=2}^{n}Y_r  \geq t-j\right)\Pr \left(Y_1=j \right)\\
&= \Pr\left(\sum_{r=1}^{n}Y_r \geq t\right).
\end{align*}

\subsection{Proof of Lemma~\ref{lmm:dominated}}

Consider the following natural coupling for $\A_1$ and $A_2$: At each time $t$, if ball $t$ goes to bin $x$ for $\A_2$, then ball $t$ goes to bin $x$ with probability $\frac{p_{x}^{\A_1}(t)}{p_{x}^{\A_2}(t)}$  and is discarded with probability $1 - \frac{p_{x}^{\A_1}(t)}{p_{x}^{\A_2}(t)}$ for $\A_1$. We show by induction that under the coupling process 
\begin{align}
B_x^{\A_1}(t) \leq B_x^{\A_2}(t) \label{eq:coupleInd}
\end{align}
holds for any $x$ and $t$, where
$B_x^{\A_i}(t)$ denotes the number of balls in bin $x$ at time $t$ (i.e., right after the $t$th ball is placed and bins are sorted).
Initially, all bins are empty: $B_x^{\A_1}(0) = B_x^{\A_2}(0) = 0$. First ball $t=1$ is placed into a bin $ x $ in $\A_2$ for some $1 \leq x \leq n$.  At the same time,  in process $\A_1$,  either ball $t$  is placed in bin $x$ or no bins receive a ball. Then, after sorting (if needed), we have $B_1^{\A_2}(1) = 1 \geq B_1^{A_1}(1) $ and $B_x^{\A_2}(1)=B_x^{\A_1}(1)=0$ for $x > 1$. Thus (\ref{eq:coupleInd}) holds when $t=1$.
Now assume that the induction hypothesis (\ref{eq:coupleInd}) holds for some $t \geq 1$.
We need to show that
\begin{align*}
B_x^{\A_1}(t+1) \leq B_x^{\A_2}(t+1). 
\end{align*}
Assume to the contrary that
\begin{align}
B_{x}^{\A_1}(t+1) > B_{x}^{\A_2}(t+1) \label{assump}
\end{align}
 for some $x$. We further assume that $x'$ is the smallest $x$ for which 
(\ref{assump}) holds. 
The two conditions (\ref{eq:coupleInd}) and  (\ref{assump}) imply that \\
\indent (A) there is a bin $z \geq x'$ that receives a ball $t+1$ in each of $\A_1$ and $\A_2$,\\
\indent (B) $B_{x'}^{\A_1}(t) = B_{x'+1}^{\A_1}(t) = ... = B_{z}^{\A_1}(t)$, and\\
\indent (C) either  $x'=1$ or $B_{x'-1}^{\A_1}(t) > B_{x'}^{\A_1}(t)$. \\
\indent (D) $B_{x'}^{\A_1}(t) = B_{x'}^{\A_2}(t)$.
\\
Combining the induction hypothesis (\ref{eq:coupleInd}), (B) and (D), we draw the following conclusion;  \\
\indent (E) $B_{x'}^{\A_2}(t) = B_{x'+1}^{\A_2}(t) = ... = B_{z}^{\A_2}(t)$.\\
If $x'>1$ and $B_{x'-1}^{\A_2}(t) > B_{x'}^{\A_2}(t)$, then we would end up with $B_{x'}^{\A_2}(t+1) = B_{x'}^{\A_2}(t)+1 = B_{x'}^{A_1}(t+1)$ which is a contradiction to the definition of $x'$. Therefore, we must have\\
\indent (F) $B_{x'-1}^{\A_2}(t) = B_{x'}^{\A_2}(t)$ if $x'>1$.\\
If $x'=1$, then $B_1^{\A_2}(t) = B_1^{\A_1}(r) = B_2^{\A_2}(t) = \ldots = B_z^{\A_1}(t)$ and hence 
$B_1^{\A_2}(r) =B _2^{\A_2}(t) = \ldots = B_z^{\A_1}(t)$, which implies $B_1^{\A_2}(t+1) = B_1^{\A_2}(t)+1 = B_1^{\A_1}(t)+1=B_1^{\A_1}(t+1)$, which contradicts to (\ref{assump}). Therefore, we must have $x'>1$.
Then, (C), (D), and (F) imply that
$B_{x'-1}^{\A_1}(t)  > B_{x'-1}^{\A_2}(t)$, which is a contradiction to the definition of $x'$.

\subsection{Proof of Lemma~\ref{pr:la}}  
The first two  properties can be derived directly by the definition of $\LA$. The last property (iii) is a consequence of Lemma~\ref{lmm:dominated}.

\subsection{Proof of Lemma~\ref{lmm:x1}}
Consider the following natural coupling  for $\SA$ and $\LA$. If the $t$th ball  is placed into a randomly chosen bin, say bin $x$ (the $x$th most loaded bin right before $t$ is placed), for $\SA$, then the $t$th ball is placed into bin $x$ only if $x \geq x_0$ in $\LA$. 
First, we show that under this coupling 
\begin{align}
\sum_{x=x_0}^{x_{1}} B_{x}^{\SA}(t) \leq \sum_{x=1}^{x_{1}} B_{x}^{\LA}(t) \label{eq:x0x1}
\end{align}
holds for any $x_1>x_0$.
Since whenever bin $x \geq x_0$ receives a ball in $\SA$, the same bin receives a ball in $\LA$, we have 
\begin{align}
\sum_{x=x_0}^{n } B_{x}^{\SA}(t) \leq \sum_{x=1}^{n}B_{x}^{\LA}(t). \label{eq:total}
\end{align}
Using Lemma~\ref{lmm:dominated}, we have
$$
B_{x}^{\SA}(t) \geq B_{x}^{\LA}(t),
$$
which implies
\begin{align}
 \sum_{x=x_1+1}^n B_{x}^{\SA}(t)  \geq \sum_{x = x_{1}+1}^{n} B_{x}^{\LA}(t).  \label{eq:tail}
\end{align}
From (\ref{eq:total}) and (\ref{eq:tail}), we obtain (\ref{eq:x0x1}).

Using the result (ii)  in Lemma~\ref{pr:la}, 
$\sum_{x=1}^{x_{1}} B_{x}^{\LA}(t) \leq x_{1} (B_{x_0}^{\LA}(t)+1).$
Since $(x_{1}-x_{0}) B_{x_{1}}^{\SA}(t) \leq\sum_{x=x_0}^{x_{1}} B_{x}^{\SA}(t)$, we obtain
$$  (x_1-x_0)B_{x_{1}}^{\SA}(t)   \leq   x_1 \left(B_{x_0}^{\LA}(t)+1\right).$$
We have shown that under the coupling we defined the above inequality holds, which implies
\begin{align*}
\Pr\left( B_{x_1}^{\SA}(t) \geq s\right) \leq \Pr\left(\frac{x_1}{x_1-x_0}  (B_{x_{0}}^{\LA}(t) +1) \geq s \right),
\end{align*}
for $s \geq 0$.


\subsection{Proof of Lemma~\ref{lmm:gamma}}

Let $S_r$ denote the set of $d$ random bins selected in the beginning of a round $r$ for $\A(k,d)$.
Fix $x \geq \gamma_*$ and let $N_x$ denote the number of bins in $S_r$ selected from the set $\{\text{bin $1$, bin $2, \ldots$, bin $x$}\}$.  Then $N_x$ has a binomial distribution with parameters $d$ and  $\frac{x}{n}$. That is, $N_x \sim B(d, \frac{x}{n})$ and $\E[N_x] = d\frac{x}{n}$. 
Let $\mathcal{E}$ denote the event that $N_x \geq d-k+2$.
Note that since $d_k \to \infty$, we have $d-k \ll d$ and hence $d-k+2 \ll d$.

Consider the following coupling for $\A$ and $\SA$.
For each round $r$,  each of $k$ balls chooses a random bin as its destination for $\SA$.  At the same time, a set $T_r$ of $d-k$ random bins is selected in $\A$ and $k$ balls are placed into $k$ least loaded bins in the set $S_r \cup T_r$, where $S_r$ is the set of $k$ random bins chosen in $\SA$.
Let $x \geq \gamma_*$. Under this coupling,
conditioning on $\mathcal{E}$,  if bin $x$  for $\A$ receives $m \geq 0$ balls in the round then bin $x$ for $\SA$ receives at most $m$ balls.
Therefore, there exists a serialization $\A_{\sigma}$ of $\A$ such that for each ball $t$
$C_{x}^{\A}(t) \geq C_{x}^{\SA}(t)$, where $C_x^{\SA}(t)$ is a Bernoulli random variable set to $1$ if and only if $t$th ball is placed into bin $x$. Therefore,
\begin{align}
\Pr\left( C_x^{\A}(t) = 1 \;|\; \mathcal{E}\right) &\geq  
\Pr\left(  C_x^{\SA}(t) = 1 \;|\; \mathcal{E}\right)  \notag \\
&\geq \Pr\left(  C_x^{\SA}(t) = 1 \;|\;  N_x  = d-x+2 \right)  \notag \\
&= 1-\left(1-\frac{1}{x}\right)^2. \label{eq:C_x}
\end{align}
In the last inequality above, we use the fact that the probability that ball $t$ goes to bin $x$ given the condition $N_x = d-x+2$  is greater than or equal to the probability that bin $x$ is selected two times randomly from the set $\{\text{bin $1$, $\ldots$, bin $x$}\}$.
By a Chernoff bound, 
\begin{align*}
\Pr \left(\mathcal{E}^c \right) &= \Pr\left(N_x \leq d-k+1\right)\\
&= \Pr \left(N_x < (1-\delta) \E[N_x] \right),  \qquad \text{where } \delta = 1 - \frac{(d-k+1)n}{xd}\\
&\leq e^{-\E[N_x]\delta^2/2}\\
&= e^{-\frac{1}{2} \left(1-\frac{d-k+1}{x\cdot d}n\right)^2\frac{x\cdot d}{n}}\\
&\leq e^{-\frac{1}{2} \left(1-\frac{d-k+1}{\gamma_*\cdot d}n\right)^2\frac{x\cdot d}{n}}\\
&\leq e^{-\frac{d}{8}\frac{x}{n}}\\
&\leq e^{-\frac{x}{n}}.
\end{align*}
From (\ref{eq:C_x}), we have
\begin{align*}
\Pr\left( C_x^{\A}(t) = 1 \right) &\geq \left(1-\left(1-\frac{1}{x}\right)^2\right) \Pr\left(\mathcal{E}\right) \\
&\geq \left(1-\left(1-\frac{1}{x}\right)^2\right)  \left(1 -e^{-\frac{x}{n}} \right)\\
&\geq \left(\frac{2}{x} -\frac{1}{x^2}\right) \frac{2x}{3n}\\
&\geq \frac{1}{n}.
\end{align*}

\section{Conclusion and Future Work}\label{sec:con}
We have examined the $(k,d)$-choice allocation process  that improves load balance, message cost, and allocation speed, making it suitable for cluster job scheduling and distributed storage. Our scheme can be viewed as a mix between the single and $d$-choice balls-into-bins models, but superior in performance. We have employed 
several new techniques such as serialization of parallel process, domination  (stronger notion than majorization) and ``unnatural'' coupling.   With our results, one can choose appropriate $k$ and $d$ to enable the  $(k,d)$-choice strategy to achieve the optimal tradeoffs between the maximum load and message cost. 

Many questions remain open regarding $(k,d)$-choice. 
First, the maximum load of a heavily loaded case, i.e., $m>n$ balls into $n$ bins, is not known for $d < 2k$.  
Considering that the fundamental differences between single- and multi-choice processes become more pronounced as $m$ grows,  it is of special interest to investigate the behavior of  $(k,d)$-choice as $m$ increases. 
The performance of $(k,d)$-choice can be further improved by adjusting the parameter $k$ dynamically in each round or by modifying the allocation policy  in the way that the less-loaded candidate bins can receive more balls regardless of how many times those bins are sampled. In $(2, 3)$-choice, for example, when $3$ bins with $0, 2$, and $3$ balls are chosen randomly, two balls are placed into the empty bin, instead of one being put into the empty and the other into the bin with $2$ balls. This adjustment may reduce the maximum load to a constant even when $k \approx d$ and $d$ is large.

\bibliographystyle{plain}
\bibliography{generalMultiChoice}

\begin{thebibliography}{10}

\bibitem{Adler95}
Micah Adler, Soumen Chakrabarti, Michael Mitzenmacher, and Lars Rasmussen.
\newblock Parallel randomized load balancing.
\newblock In {\em Symposium on Theory of Computing. ACM}, pages 119--130, 1995.

\bibitem{ABKU94}
Yossi Azar, Andrei~Z. Broder, Anna~R. Karlin, and Eli Upfal.
\newblock Balanced allocations.
\newblock {\em SIAM Journal on Computing}, 29(1):180--200, September 1999.

\bibitem{BCEFN12}
P.~Berenbrink, A.~Czumaj, M.~Englert, T.~Friedetzky, and L.~Nagel.
\newblock Multiple-choice balanced allocation in (almost) parallel.
\newblock In {\em Proceedings of the International Workshop on Randomization
  and Computation (RANDOM 2012)}, pages 411--422, 2012.

\bibitem{Cloud13}
Petra Berenbrink, Andre Brinkmann, Tom Friedetzky, Dirk Meister, and Lars~Nagel
  Dirk.
\newblock Distributing storage in cloud environments.
\newblock In {\em Proceedings of High-Performance Grid and Cloud Computing
  Workshop (workshop of IPDPS)}, 2013.

\bibitem{BCSV00}
Petra Berenbrink, Artur Czumaj, Angelika Steger, and Berthold V{\"{o}}cking.
\newblock Balanced allocations: The heavily loaded case.
\newblock {\em {SIAM} Journal on Computing}, 35(6):1350--1385, 2006.

\bibitem{BSS13}
Petra Berenbrink, Kamyar Khodamoradi, Thomas Sauerwald, and Alexandre Stauffer.
\newblock Balls-into-bins with nearly optimal load distribution.
\newblock In {\em Proceedings of the 25th Symposium on Parallelism in
  Algorithms and Architectures (SPAA)}, pages 326--335, 2013.

\bibitem{Czumaj}
A.~Czumaj and V.~Stemann.
\newblock Randomized allocation processes.
\newblock {\em Random Structures and Algorithms}, 18(4):297--331, June 2001.

\bibitem{Dubhashi96}
Devdatt Dubhashi and Desh Ranjan.
\newblock Balls and bins: A study in negative dependence.
\newblock {\em Random structures \& algorithms}, 13:99--124, 1996.

\bibitem{G08}
P.~Brighten Godfrey.
\newblock Balls and bins with structure: balanced allocations on hypergraphs.
\newblock In {\em ACM-SIAM Symposium on Discrete Algorithms}, pages 511--517,
  2008.

\bibitem{Lenzen}
Christoph Lenzen and Roger Wattenhofer.
\newblock Tight bounds for parallel randomized load balancing: extended
  abstract.
\newblock In {\em STOC '11}, pages 11--20, 2011.

\bibitem{Power96}
Michael Mitzenmacher.
\newblock {\em The Power of Two Choices in Randomized Load Balancing}.
\newblock PhD thesis, University of California, Berkeley, 1996.

\bibitem{SPARROW}
Kay Ousterhout, Patrick Wendell, Matei Zaharia, and Ion Stoica.
\newblock Sparrow: Distributed, low latency scheduling.
\newblock In {\em Proceedings of the Twenty-Fourth ACM Symposium on Operating
  Systems Principles}, SOSP '13, pages 69--84, New York, NY, USA, 2013. ACM.

\bibitem{PARK11}
Gahyun Park.
\newblock Brief announcement: A generalization of multiple choice
  balls-into-bins.
\newblock In {\em PODC '11}, pages 297--298, 2011.

\bibitem{PTW10}
Yuval Peres, Kunal Talwar, and Udi Wieder.
\newblock The $(1+\beta)$-choice process and weighted balls into bins.
\newblock In {\em ACM-SIAM Symposium on Discrete Algorithms}, pages 1613--1619,
  2010.

\bibitem{RS98}
Martin Raab and Angelika Steger.
\newblock Balls into bins - a simple and tight analysis, 1998.

\bibitem{Stemann96}
Volker Stemann.
\newblock Parallel balanced allocations.
\newblock In {\em Proceedings of the Eighth Annual ACM Symposium on Parallel
  Algorithms and Architectures}, SPAA '96, pages 261--269, New York, NY, USA,
  1996. ACM.

\bibitem{Talwar07}
Kunal Talwar and Udi Wieder.
\newblock Balanced allocations: the weighted case.
\newblock In {\em Proceedings of the thirty-ninth annual ACM Symposium on
  Theory of Computing}, STOC '07, pages 256--265, 2007.

\bibitem{Talwar13}
Kunal Talwar and Udi Wieder.
\newblock Balanced allocations: {A} simple proof for the heavily loaded case.
\newblock In {\em Automata, Languages, and Programming - 41st International
  Colloquium, {ICALP} 2014, Copenhagen, Denmark, July 8-11, 2014, Proceedings,
  Part {I}}, pages 979--990, 2014.

\bibitem{V99}
Berthold V\"{o}cking.
\newblock How asymmetry helps load balancing.
\newblock {\em Journal of ACM}, 50(4):568--589, July 2003.

\end{thebibliography}

\end{document}